\documentclass[10pt,journal]{IEEEtran}

\usepackage{amssymb}
\usepackage{amsmath}
\usepackage{cite}
\usepackage{url}
\usepackage{xcolor}
\usepackage{cite,graphicx,amsmath,amssymb}
\usepackage{fancyhdr}
\usepackage{mdwmath}
\usepackage{mdwtab}
\usepackage{caption}
\usepackage{amsthm}
\usepackage{setspace}
\usepackage{hyperref}
\usepackage{algorithm}
\usepackage{algorithmic}
\usepackage{multirow}
\usepackage{makecell}
\usepackage{mathtools}
\usepackage{subcaption}
\usepackage{mathrsfs}
\usepackage{bm}
\usepackage{pifont}

\usepackage{booktabs}
\usepackage{multirow}
\usepackage{siunitx}
\usepackage{balance}

\hypersetup{colorlinks=true,
linkcolor=blue,
citecolor=blue,      
urlcolor=black,
    }

\graphicspath{{./src/img/}}

\newtheorem{theorem}{Theorem}

\newtheorem{lemma}{Lemma}

\newtheorem{corollary}{Corollary}

\newtheorem{proposition}{Proposition}

\allowdisplaybreaks
\title{Beamforming Optimization for Continuous Aperture Array (CAPA)-based Communications}

\author{
        Zhaolin Wang,~\IEEEmembership{Member,~IEEE,}
        Chongjun Ouyang,~\IEEEmembership{Member,~IEEE,} \\
        and Yuanwei Liu,~\IEEEmembership{Fellow,~IEEE}
\thanks{The work of Chongjun Ouyang was supported in part by the Marie Skłodowska-Curie Actions (MSCA) Postdoctoral Fellowship and in part by the U.K. Engineering and Physical Sciences Research Council (EPSRC) under Grant EP/Z003091/1. An earlier version of this paper will be presented in part at the 2025 IEEE International Conference on Communications, Montreal, Canada.}
\thanks{Zhaolin Wang and Chongjun Ouyang are with the School of Electronic Engineering and Computer Science, Queen Mary University of London, London E1 4NS, U.K. (e-mail: \{zhaolin.wang, c.ouyang\}@qmul.ac.uk).}
\thanks{Yuanwei Liu is with the Department of Electrical and Electronic Engineering, The University of Hong Kong, Hong Kong, and also with the Department of Electronic Engineering, Kyung Hee University, Yongin-si, Gyeonggi-do 17104, South Korea (e-mail: yuanwei@hku.hk).}
\vspace{-0.3cm}
}

\begin{document}

\maketitle
\begin{abstract}

    The beamforming optimization in continuous aperture array (CAPA)-based multi-user communications is studied. In contrast to conventional spatially discrete antenna arrays, CAPAs can exploit the full spatial degrees of freedom (DoFs) by emitting information-bearing electromagnetic (EM) waves through continuous source current distributed across the aperture. Nevertheless, such an operation renders the beamforming optimization problem as a non-convex integral-based functional programming problem, which is challenging for conventional discrete optimization methods. A couple of low-complexity approaches are proposed to solve the functional programming problem. 1) \emph{Calculus of variations (CoV)-based approach:} Closed-form structure of the optimal continuous source patterns are derived based on CoV, inspiring a low-complexity integral-free iterative algorithm for solving the functional programming problem. 2) \emph{Correlation-based zero-forcing (Corr-ZF) approach:} Closed-form ZF source current patterns that completely eliminate the inter-user interference are derived based on the channel correlations. By using these patterns, the original functional programming problem is transformed to a simple power allocation problem, which can be solved using the classical water-filling approach with reduced complexity. Our numerical results validate the effectiveness of the proposed designs and reveal that: \emph{i)} compared to the state-of-the-art Fourier-based discretization approach, the proposed CoV-based approach not only improves communication performance but also reduces computational complexity by up to hundreds of times for large CAPA apertures and high frequencies, and \emph{ii)} the proposed Corr-ZF approach achieves asymptotically optimal performance compared to the CoV-based approach.
    
\end{abstract}


\section{Introduction} \label{sec:intro}

\IEEEPARstart{S}INCE the advent of 3G, multiple-input multiple-output (MIMO) techniques have been an essential element of wireless communication standards in enhancing communication capacity \cite{andrews2014will}. By equipping transceivers with multiple antennas, MIMO leverages the spatial degrees of freedom (DoFs) for communications, significantly multiplying spectral efficiency. While the spatial DoFs and capacity of MIMO communications grow monotonically with the number of antennas, they are inherently limited by the physical constraints of antenna deployment. An effective way to practically approach the optimal spatial DoFs and capacity limits is by integrating a large number of antennas into a limited surface area, which has driven the evolution toward Massive MIMO in 5G \cite{andrews2014will} and Gigantic MIMO in 6G \cite{bjornson2024enabling}, with the aim of integrating hundreds or even thousands of antennas at transceivers. To achieve these goals, many state-of-the-art antenna array architectures have been proposed, such as holographic MIMO \cite{ huang2020holographic, 9110848, 10232975}, large intelligent surface \cite{hu2018beyond, dardari2020communicating}, and dynamic metasurface antennas \cite{shlezinger2021dynamic}.  The ultimate goal of these array architectures is to form a (approximately) continuous electromagnetic (EM) aperture,  referred to as \emph{continuous aperture array (CAPA)}.  

CAPA can be more attractive than conventional MIMO with spatially discrete antennas for achieving high spatial DoFs and capacity in wireless communications. First, unlike conventional MIMO, where the spatial DoFs are limited by the number of discrete antennas, the continuous surface of CAPA eliminates this limitation, allowing it to fully exploit the spatial DoFs provided by the aperture. Second, in conventional MIMO systems, each antenna requires dedicated hardware. For example, in fully digital arrays, each antenna needs its own radio frequency (RF) ports, which can pose significant implementation and energy challenges for massive or gigantic MIMO systems \cite{7400949}. Even in hybrid analog-digital arrays, each antenna requires a dedicated analog phase shifter, and the total number of RF ports must be at least twice the number of signals to spatially multiplex in order to achieve optimal performance \cite{7389996}. In contrast, CAPA only requires an equal number of RF ports to the number of spatially multiplexed signals \cite{bjornson2019massive}, making it a more efficient and scalable solution.

\subsection{Prior Works}
The modelling, performance analysis, and system optimization of CAPA are usually studied from the perspective of EM theory to address the challenges posed by its continuous EM radiating surface. Specifically, the transmit signal is represented by information-bearing sinusoidal source currents distributed across the aperture of CAPA, which generate EM waves that propagate toward the communication users. The users then receive these EM waves and decode the desired information from them. This modelling approach has recently been referred to as EM information theory. Based on this approach, the fundamental communication performance between two CAPAs has been widely studied since the 1980s. In particular, the authors of \cite{bucci1989degrees} investigated the DoFs of scattered EM fields, revealing their relation to the Nyquist number associated with the observation interval. From the EM formulation, the authors \cite{miller2000communicating} proved that the communication DoFs are proportional to the volumes of the transmit and receive CAPAs by solving eigenfunction problems. As a further advance, accurate analytical expressions for the communication DoFs between two CAPAs were derived in \cite{dardari2020communicating}, showing that the DoFs are greater than one, even in line-of-sight (LoS) scenarios. This study was extended to near-field scenarios in \cite{decarli2021communication}, where a practical method based on the multi-focusing capability of large CAPA was proposed to approximate the optimal communication DoFs. In \cite{9848802}, the authors developed a spatial bandwidth approach to analyze the DoFs between two linear CAPAs, deriving interpretable closed-form approximations for the DoFs for three orthogonal receiving directions. In a separate study, the authors of \cite{10262267} explored the effective DoFs between two linear CAPAs.

Channel capacity is another important issue for CAPA systems and has attracted significant research attention. For example, the authors of \cite{jensen2008capacity} proposed a general EM framework for characterizing the capacity between two CAPAs using the eigenfunction method, which revealed the bounded capacity for finite transmit power and provided solutions to mitigate supergain effects through practical constraints. The impact of physical losses in CAPAs on channel capacity and DoFs was investigated in \cite{jeon2017capacity}, showing how capacity is influenced by radiation efficiency and the Q factor. In \cite{8585146}, the authors applied Kolmogorov information theory to characterize the channel capacity between two CAPAs in the presence of asymmetric DoFs in the time and space domains. Furthermore, the authors of \cite{10303285} compared the channel capacity of CAPA and conventional discrete MIMO systems under equal power allocation among data streams, showing that conventional discrete MIMO can approach CAPA performance when an infinite number of antennas without mutual coupling are deployed. As a further advancement, the authors of \cite{wan2023mutual} proposed an effective numerical calculation scheme for channel capacity between two CAPAs over random fields. In a separate study, \cite{10012689} developed a general method to compute channel capacity between two CAPAs with arbitrary radiating surfaces. Inspired by orthogonal frequency division multiplexing, the authors of \cite{9906802} introduced an effective wavenumber-division multiplexing transmission scheme to approach the channel capacity between two CAPAs under LoS conditions.

The aforementioned works all focused on the communication performance between a single pair of transceivers. There have also been several initial works for the multi-user scenarios. In particular, the authors of \cite{hu2018beyond} studied an uplink multi-user CAPA system under LoS conditions, where uplink capacity was analyzed based on the matched-filtering operation. Under the same condition, the signal-to-interference-plus-noise ratio (SINR) in uplink multi-user CAPA systems was analyzed in \cite{10669060}, upon which an adaptive interference mitigation method was devised. As a further advance, the authors of \cite{zhao2024continuous} investigated the uplink multi-user capacity with CAPAs equipped at both transceivers, followed by an analysis of the downlink multi-user capacity based on the uplink-downlink duality. Some studies have approached multi-user CAPA systems from an optimization perspective, given the limited generalization and scalability of existing methods. For example, beamforming optimization in multi-user downlink CAPA systems was first explored in \cite{zhang2023pattern}, where a Fourier-based approach was proposed to maximize the sum rate. The study demonstrated that optimized transmit beamformers significantly enhance the advantages of CAPA over conventional MIMO, while simple matched-filtering can result in substantial performance loss. Building on this Fourier-based approach, beamforming optimization for multi-user uplink CAPA systems was further investigated in \cite{10612761}.

\subsection{Motivation and Contributions}
Beamforming optimization has long been recognized as a key technique for maximizing communication performance in conventional MIMO systems \cite{5447076}. By strategically adjusting the signal phase and amplitude at each antenna, beamforming enhances signal strength and mitigates interference, leading to significant gains in communication capacity. Recent research has extended these advantages to CAPA systems \cite{zhang2023pattern, 10612761}, demonstrating that beamforming optimization can further unlock the potential of CAPA by fully exploiting its continuous aperture for enhanced spatial DoFs. However, in CAPA systems, the beamformers to be optimized are no longer discrete vectors or matrices but a set of spatially continuous source current patterns distributed across the aperture. This makes it challenging to directly apply conventional matrix-based optimization methods designed for discrete systems. The Fourier-based approach is the most common approach for addressing this challenge to date \cite{zhang2023pattern, 10612761}. This approach approximates the continuous source current patterns using a finite number of Fourier basis functions, transforming the optimization problem from continuous to discrete. Therefore, the conventional matrix-based optimization methods can still be applied. However, the Fourier-based approach faces two significant challenges. First, the number of Fourier basis functions required to accurately approximate the original continuous patterns grows exponentially with the CAPA aperture size and signal frequency. This results in a high-dimensional optimization problem, potentially leading to prohibitive computational complexity. Second, because the approach optimizes discrete, approximated source current patterns, it becomes difficult to evaluate the true performance limits of CAPA systems.

To the best of our knowledge, no existing work has attempted to directly optimize the continuous source current patterns to maximize communication capacity in CAPA systems. This raises an interesting and fundamental question: \emph{if the continuous source current patterns are directly optimized, what performance limits can CAPA achieve, and how might computational complexity be affected—whether increased or reduced?} To answer this question, we propose an iterative CAPA beamforming approach based on the \emph{calculus of variations (CoV)}. In contrast to the state-of-the-art Fourier-based approach, the proposed CoV-based approach can directly optimize the continuous source current patterns without any approximations. It is interesting to find that directly optimizing the continuous source current patterns not only improves performance but also significantly reduces computational complexity, which becomes independent of the CAPA aperture size and signal frequency. To further reduce complexity, we also introduce a new \emph{correlation-based zero-forcing (Corr-ZF)} approach for CAPA beamforming, deriving closed-form pseudoinverse-free ZF source current patterns that achieve asymptotically optimal performance. The main contributions of this paper can be summarized as follows:
\begin{itemize}
    \item We investigate the CAPA beamforming optimization problem for maximizing the weighted sum rate (WSR) of multiple downlink communication users, where the transmit beamformers are modelled as continuous source current patterns based on EM theory.
    \item We propose a CoV-based approach to directly optimize the continuous source current patterns without any approximations. Based on the idea of CoV, we derive the structure of the optimal continuous source current patterns, leading to a low-complexity integral-free block coordinate descent (BCD) algorithm.
    \item We further propose a Corr-ZF approach for CAPA beamforming to reduce the computational complexity. We derive the closed-form pseudoinverse-free ZF continuous source current patterns based on channel correlations to completely eliminate the inter-user interference. Then, we exploit the classical water-filling method to address the remaining power allocation problem among the users.
    \item We provide comprehensive numerical results to validate the effectiveness and efficiency of the proposed approaches. The results demonstrate that, compared to the state-of-the-art Fourier-based approaches, the proposed CoV-based approach achieves superior WSR performance under various system settings while significantly reducing computational complexity by hundreds of times for large apertures and high frequencies. Additionally, the proposed Corr-ZF approach is shown to achieve asymptotically optimal WSR performance compared to the proposed CoV-based approach.
\end{itemize}

\subsection{Organization and Notations}
The rest of this paper is organized as follows. Section \ref{sec:model} presents the system model and the formulation of the WSR problem for CAPA beamforming. Sections \ref{sec:CoV} and \ref{sec:ZF} elaborate on the proposed CoV-based and Corr-ZF approaches for CAPA beamforming, respectively. Section \ref{sec:Fourier_method} compares the proposed approaches with the state-of-the-art Fourier-based approach. Section \ref{sec:results} provides numerical results comparing different approaches under various system settings. Finally, Section \ref{sec:conclusion} concludes this paper.

\emph{Notations:} Scalars, vectors/matrices, and Euclidean subspaces are represented by regular, boldface, and calligraphic letters, respectively. The set of complex, real, and positive real numbers are denoted by $\mathbb{C}$, $\mathbb{R}$, and $\mathbb{R}_+$, respectively. The set of symmetric positive semidefinite matrices of dimension $N \times N$ is denoted by $\mathbb{S}_+^{N \times N}$. The inverse, conjugate, transpose, conjugate transpose, and trace operators are denoted by $(\cdot)^{-1}$, $(\cdot)^*$, $(\cdot)^T$, $(\cdot)^H$, and $\mathrm{tr}(\cdot)$, respectively. The Lebesgue measure of a Euclidean subspace $\mathcal{S}$ is denoted by $|\mathcal{S}|$. The absolute value and Euclidean norm are denoted by $|\cdot|$ and $\|\cdot\|$, respectively. The ceiling operator is denoted by $\lceil \cdot \rceil$. The real part of a complex number is demoted by $\Re \{\cdot\}$. An identity matrix of dimension $N \times N$ is denoted by $\mathbf{I}_N$. The big-O notation is denoted by $O(\cdot)$. 
\section{System Model and Problem Formulation} \label{sec:model}
\begin{figure}[t!]
    \centering
    \includegraphics[width=0.45\textwidth]{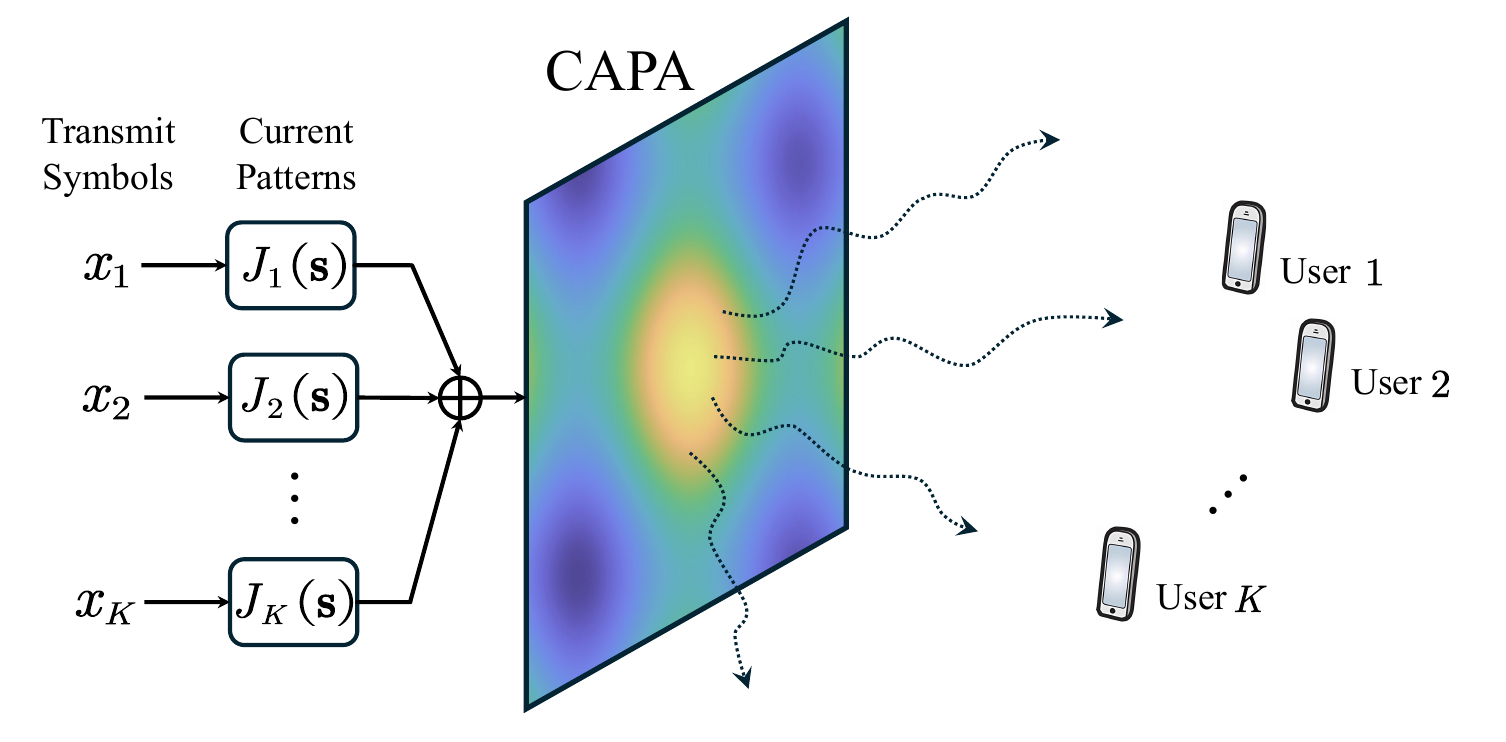}
    \caption{Illustration of a CAPA-based multi-user communications.}
    \label{fig_system_model}
\end{figure}

As illustrated in Fig. \ref{fig_system_model}, we study a CAPA-based multi-user downlink communication system, where a CAPA transmitter is equipped at the base station to serve $K$ single-antenna communication users. We model the system from the EM theory perspective as elaborated in the following.

\subsection{Transmit Signal}
We consider a CAPA transmitter with a continuous surface $\mathcal{S}_{\mathrm{T}}$ with an area of $A_{\mathrm{T}} = |\mathcal{S}_{\mathrm{T}}|$, which contains sinusoidal source currents to emit EM waves for wireless communications. Let $\mathbf{J}(\mathbf{s}, \omega) \in \mathbb{C}^{3 \times 1}$ represent the Fourier transform of the source current density at point $\mathbf{s} = [s_x, s_y, s_z]^T \in \mathcal{S}_{\mathrm{T}}$, where $\omega = 2 \pi f / c = 2 \pi / \lambda$ denotes the angular frequency, $f$ is the signal frequency, and $\lambda$ is the signal wavelength. For brevity, and without loss of generality, we focus on a narrowband single-carrier communication system, where the explicit dependence of the source current on $\omega$ can be omitted. Hence, the source current is denoted as $\mathbf{J}(\mathbf{s})$. The wideband multi-carrier system can be addressed by treating each carrier frequency separately. Additionally, the source current can be decomposed into orthogonal components along different directions as follows:
\begin{equation}
    \mathbf{J}(\mathbf{s}) = J_x(\mathbf{s}) \hat{\mathbf{u}}_x + J_y (\mathbf{s}) \hat{\mathbf{u}}_y + J_z(\mathbf{s}) \hat{\mathbf{u}}_z,
\end{equation}
where $\hat{\mathbf{u}}_x \in \mathbb{R}^{3 \times 1}$, $\hat{\mathbf{u}}_y \in \mathbb{R}^{3 \times 1}$, and $\hat{\mathbf{u}}_z \in \mathbb{R}^{3 \times 1}$ are unit vectors along the $x$-, $y$-, and $z$-axes, respectively. In this study, we consider the case of a vertically polarized transmitter, where only the $y$-component of the source current is excited. In this case, the source current becomes
\begin{equation} \label{y_source_current}
    \mathbf{J}(\mathbf{s}) = J(\mathbf{s}) \hat{\mathbf{u}}_y,
\end{equation}
where we define $J(\mathbf{s}) := J_y(\mathbf{s})$ to simplify the notation. To convey information to $K$ users, the scalar source current $J(\mathbf{s})$ is a linear superposition of $K$ information-bearing source currents, expressed as
\begin{equation} \label{y_source_current_separate}
    J(\mathbf{s}) = \sum_{k=1}^K J_k(\mathbf{s}) x_k,
\end{equation}
where $J_k(\mathbf{s}) \in \mathbb{C}$ and $x_k \in \mathbb{C}$ represent the source current pattern and the communication symbol for the $k$-th user, respectively. The source current pattern $J_k(\mathbf{s})$ is analogous to the transmit beamformer in conventional spatially discrete antenna systems. The communication symbols are assumed to be independent and have unit power, satisfying $\mathbb{E}\{\mathbf{x} \mathbf{x}^H\} = \mathbf{I}_K$, where $\mathbf{x} = [x_1, \dots, x_K]^T$.

\subsection{EM Channel and Receive Signal}
Let $\mathbf{r}_k \in \mathbb{R}^{3 \times 1}$ denote the location of the $k$-th user. From Maxwell's equations, the electric filed at the location $\mathbf{r}_k$ generated by the source current $\mathbf{J}(\mathbf{s})$ in a homogeneous medium is given by \cite{1386525, dardari2020communicating}
\begin{equation}
    \label{E_field}
    \mathbf{E}_k = \int_{\mathcal{S}_{\mathrm{T}}} \mathbf{G}(\mathbf{r}_k, \mathbf{s}) \mathbf{J}(\mathbf{s}) d \mathbf{s} \in \mathbb{C}^{3 \times 1}.
\end{equation}  
For line-of-sight scenarios, the integral kernel $\mathbf{G}(\mathbf{r}, \mathbf{s}) \in \mathbb{C}^{3 \times 1}$ is often referred to as the Green's function. In the region where the EM field has settled into normal radiation, $\mathbf{G}(\mathbf{r}, \mathbf{s})$ can be expressed as\footnote{In \eqref{Green_function}, the higher-order terms associated with the reactive near-field effect are omitted as their impact on system performance is negligible \cite{10614327}.}
\begin{align} \label{Green_function}
    \mathbf{G}(\mathbf{r}, \mathbf{s}) = - \frac{j \eta e^{- j \frac{2\pi}{\lambda} \|\mathbf{r} - \mathbf{s}\|}}{2 \lambda \|\mathbf{r} - \mathbf{s}\|} \left( \mathbf{I}_3 - \frac{(\mathbf{r}-\mathbf{s})(\mathbf{r} - \mathbf{s})^T}{ \|\mathbf{r} - \mathbf{s}\|^2} \right),
\end{align}
where $\eta$ denotes the intrinsic impedance. In rich-scattering environments, $\mathbf{G}(\mathbf{r}, \mathbf{s})$ can be modeled as a stochastic process. In the rest of this paper, we use the line-of-sight channel as an example. However, it is important to note that our proposed approaches are not limited to any specific channel type.

To capture the full three-dimensional electric field $\mathbf{E}_k$, an ideal tri-polarization receiver would be required at each user $k$, which is challenging to implement. Therefore, in this study, we assume a more practical uni-polarized antenna at each user $k$  with a polarization direction $\hat{\mathbf{u}}_k \in \mathbb{R}^{3 \times 1}$, where $\|\hat{\mathbf{u}}_k\| = 1$. This implies that user $k$ captures only the component of the electric field $\mathbf{E}_k$ along the direction $\hat{\mathbf{u}}_k$. Therefore, the noisy electric field captured by the receiver at user $k$ is
\begin{equation} \label{y_E_field}
    E_k = \hat{\mathbf{u}}_k^T \mathbf{E}_k + n_k = \int_{\mathcal{S}_{\mathrm{T}}} \hat{\mathbf{u}}_k^T \mathbf{G}(\mathbf{r}_k, \mathbf{s}) \mathbf{J}(\mathbf{s}) d \mathbf{s} + n_k,
\end{equation}
where $n_k \in \mathbb{C}$ is the EM noise and can be modelled as independent white Gaussian variables with zero mean and variance $\sigma_k^2$, i.e., $n_k \sim \mathcal{CN}(0, \sigma_k^2)$ \cite{zhang2023pattern}.    
Substituting \eqref{y_source_current} and \eqref{y_source_current_separate} into \eqref{y_E_field} yields
\begin{align} \label{receive_signal}
    E_k = &\underbrace{\int_{\mathcal{S}_{\mathrm{T}}} H_k(\mathbf{s}) J_k(\mathbf{s}) x_k d \mathbf{s}}_{\text{desired signal}} \nonumber \\
    & \hspace{1cm}+ \underbrace{\sum_{i=1, i \neq k}^K \int_{\mathcal{S}_{\mathrm{T}}} H_k(\mathbf{s}) J_i(\mathbf{s}) x_i d \mathbf{s}}_{\text{inter-user interference}} + n_k,
\end{align}
where $H_k(\mathbf{s})$ denotes the continuous EM channel for user $k$,  given by 
\begin{equation}
    H_k(\mathbf{s}) = \hat{\mathbf{u}}_k^T \mathbf{G}(\mathbf{r}_k, \mathbf{s}) \hat{\mathbf{u}}_y.
\end{equation} 

\subsection{Achievable Communication Rate}
According to Shannon's theorem, the signal-to-interference-plus-noise ratio (SINR) needs to be characterized for evaluating the achievable communication rate. Let $\varepsilon_k$ denote the absorption efficiency of the receiver at user $k$. The expected power density received at user $k$ can be calculated from \eqref{receive_signal} as 
\begin{align}
    P_k = &\mathbb{E} \left\{\frac{\varepsilon_k}{2\eta}  |E_k|^2 \right\} \nonumber \\
    = & \frac{\varepsilon_k}{2 \eta} \sum_{i=1}^K \sum_{j=1}^K \mathbb{E} \left\{x_i x_j^*\right\} \left( \int_{\mathcal{S}_{\mathrm{T}}} H_k(\mathbf{s}) J_i(\mathbf{s}) d \mathbf{s} \right) \nonumber \\
    & \hspace{2.5cm}\times\left( \int_{\mathcal{S}_{\mathrm{T}}} H_k^*(\mathbf{s}) J_j^*(\mathbf{s}) d \mathbf{s} \right) + \frac{\varepsilon_k}{2\eta}\sigma_k^2\nonumber \\
    = &\frac{\varepsilon_k}{2 \eta} \sum_{i=1}^K  \left| \int_{\mathcal{S}_{\mathrm{T}}} H_k(\mathbf{s}) J_i(\mathbf{s}) d \mathbf{s} \right|^2 + \frac{\varepsilon_k}{2\eta}\sigma_k^2,
\end{align}
where the last step is obtained by $\mathbb{E}\{\mathbf{x} \mathbf{x}^H\} = \mathbf{I}_K$. Therefore, the SINR for decoding the desired signal at user $k$ is given by
\begin{equation} \label{SINR_original}
    \gamma_k = \frac{\left| \int_{\mathcal{S}_{\mathrm{T}}} H_k(\mathbf{s}) J_k(\mathbf{s}) d \mathbf{s} \right|^2 }{\sum_{i=1,i\neq k}^K \left| \int_{\mathcal{S}_{\mathrm{T}}} H_k(\mathbf{s}) J_i(\mathbf{s}) d \mathbf{s} \right|^2 + \sigma_k^2}
\end{equation}
The achievable rate is thus given by $R_k = \log_2(1 + \gamma_k)$. 

\subsection{Problem Formulation}
We aim to optimize the source current pattern $\{J_k(\mathbf{s})\}_{k=1}^K$ to maximize the WSR of all users. The resulting optimization problem is given by 
\begin{subequations} \label{problem_se_max}
    \begin{align}
        \max_{\{ J_k(\mathbf{s}) \}_{k=1}^K} \quad &\sum_{k=1}^K \alpha_k \log_2 \left( 1 + \gamma_k \right) \\
        \label{constraint_power}
        \mathrm{s.t.} \quad & \sum_{k=1}^K \int_{\mathcal{S}_{\mathrm{T}}} | J_k(\mathbf{s}) |^2 d \mathbf{s} \le P_{\mathrm{T}}.
    \end{align}
\end{subequations}
Here, $\alpha_k$ is the weight specified for user $k$, which can be determined according to the fairness and quality of service (QoS) requirements. Constraint \eqref{constraint_power} limits the transmit power of the CAPA transmitter \cite[Lemma 1]{zhang2023pattern}. Problem \eqref{problem_se_max} is essentially a non-convex \emph{functional programming} problem\footnote{A functional is a specific type of function that takes a function as its input and produces a scalar, namely a \emph{function of function}.}, which is generally challenging to solve for the following reasons. First, the optimization variable $J_k(\mathbf{s})$ is not a finite-size vector or matrix, but a continuous function that can be viewed as an infinite-dimensional vector. Second, both the objective function and constraints are defined in terms of integrals. This kind of problem is typically solved by using commercial EM simulation software, which may result in extremely high computational complexity. The Fourier-based approach \cite{zhang2023pattern}, as will be detailed in Section \ref{sec:Fourier_method}, is a state-of-the-art alternative to address this problem recently. However, the complexity of this approach is still high, as it requires a large number of basis functions to approximate the function $J_k(\mathbf{s})$. Additionally, this approximation makes it difficult to guarantee the optimality of the obtained solution. To address the above challenges, we propose two approaches: the CoV-based approach and the Corr-ZF approach. These approaches can directly optimize the continuous function $J_k(\mathbf{s})$ without approximation, while maintaining low computational complexity.

\section{CoV-based Approach} \label{sec:CoV}

In this section, we develop a CoV-based approach that directly optimizes the source current patterns $\{\bar{J}_k(\mathbf{s})\}_{k=1}^K$, i.e., the continuous beamformers, to maximize the WSR without using any approximations.

\subsection{Problem Reformulation}
To streamline the optimization process, we first transform problem \eqref{problem_se_max} into an unconstrained optimization problem utilizing the following lemmas.

\begin{lemma} \label{lemma_equal_power}
    \normalfont
    \emph{(Equality Power Constraint)}
    The optimal solution to problem \eqref{problem_se_max} satisfies the power constraint with equality:
    \begin{equation} \label{equal_power}
        \sum_{k=1}^K \int_{\mathcal{S}_{\mathrm{T}}} | J_k(\mathbf{s}) |^2 d \mathbf{s} = P_{\mathrm{T}}.
    \end{equation}
\end{lemma}

\begin{IEEEproof}
    Please refer to Appendix \ref{lemma_equal_power_proof}.
\end{IEEEproof}

\begin{lemma} \label{lemma_power}
    \normalfont
    \emph{(Unconstrained Equivalence Problem)}
    Let $\{\bar{J}_k(\mathbf{s})\}_{k=1}^K$ denote an optimal solution to the following functional maximization problem:
    \begin{align} \label{problem_no_power}
        \max_{\{ J_k(\mathbf{s}) \}_{k=1}^K} \quad \sum_{k=1}^K \alpha_k \log_2 \left( 1 + \bar{\gamma}_k \right),
    \end{align}
    where 
    \begin{equation}
        \bar{\gamma}_k = \frac{\left| \int_{\mathcal{S}_{\mathrm{T}}} H_k(\mathbf{s}) J_k(\mathbf{s}) d \mathbf{s} \right|^2 }{ \left(\splitfrac{\sum_{i=1,i\neq k}^K \left| \int_{\mathcal{S}_{\mathrm{T}}} H_k(\mathbf{s}) J_i(\mathbf{s}) d \mathbf{s} \right|^2}{ + \frac{\sigma_k^2}{P_{\mathrm{T}}} \sum_{i=1}^K \int_{\mathcal{S}_{\mathrm{T}}} | J_i(\mathbf{s}) |^2 d \mathbf{s}}\right)}.
    \end{equation}
    An optimal solution to the problem in \eqref{problem_se_max} can then be expressed as
    \begin{equation} \label{scaled_solution}
        J_k(\mathbf{s}) = \sqrt{ \frac{P_{\mathrm{T}}}{\sum_{k=1}^K \int_{\mathcal{S}_{\mathrm{T}}} | \bar{J}_k(\mathbf{s}) |^2 d \mathbf{s}} } \bar{J}_k(\mathbf{s}).
    \end{equation}
\end{lemma}

\begin{IEEEproof}
    It can be readily shown that the solution in \eqref{scaled_solution} satisfies the equality power constraint \eqref{equal_power} and ensures that the objective function of problem \eqref{problem_se_max} always attains the same value as the objective function of problem \eqref{problem_no_power}. Therefore, according to the conclusion attained in \textbf{Lemma \ref{lemma_equal_power}} and the fact that $\bar{J}_k(\mathbf{s})$ maximizes the objective function of problem \eqref{problem_no_power}, the solution $J_k(\mathbf{s})$ in \eqref{scaled_solution} must maximize the objective function of problem \eqref{problem_se_max}. This completes the proof.
\end{IEEEproof}

\begin{lemma} \label{lemma_FP}
    \normalfont 
    \emph{(Non-fractional Equivalence Problem)}
    Problem \eqref{problem_no_power} can be equivalently reformulated as
    \begin{align} \label{problem_no_power_no_fra}
        \max_{\{\mu_k, \lambda_k, J_k(\mathbf{s}) \}_{k=1}^K} \quad &\sum_{k=1}^K \alpha_k \Bigg( 2\mu_k \Re \left\{ \lambda_k^* \int_{\mathcal{S}_{\mathrm{T}}} H_k(\mathbf{s}) J_k(\mathbf{s}) d \mathbf{s} \right\} \nonumber \\
        &- |\lambda_k^2| \Bigg(\sum_{i=1}^K \left| \int_{\mathcal{S}_{\mathrm{T}}} H_k(\mathbf{s}) J_i(\mathbf{s}) d \mathbf{s} \right|^2 \nonumber \\ &+ \frac{\sigma_k^2}{P_{\mathrm{T}}} \sum_{i=1}^K \int_{\mathcal{S}_{\mathrm{T}}} | J_i(\mathbf{s}) |^2 d \mathbf{s} \Bigg)  \Bigg),
    \end{align}
    where $\{\mu_k\}_{k=1}^K$ and $\{\lambda_k\}_{k=1}^K$ are auxiliary variables.  
\end{lemma}

\begin{IEEEproof}
    This lemma can be proved using quadratic transform \cite[Theorem 2]{shen2018fractional} and Lagrangian dual transform \cite[Theorem 3]{shen2018fractional2}. Hence, we omit the details here.
\end{IEEEproof}

Based on \textbf{Lemmas \ref{lemma_equal_power}}-\textbf{\ref{lemma_FP}}, the optimal solution to problem \eqref{problem_no_power_no_fra} must also be the optimal solution to the original problem \eqref{problem_se_max}. In the following, we propose a BCD-CoV algorithm to solve problem \eqref{problem_no_power_no_fra}.

\subsection{BCD-CoV Algorithm}

In problem \eqref{problem_no_power_no_fra}, the constraints and the fraction in the objective function have been safely removed. The optimization variables are also no longer coupled. This observation motivates the use of BCD to solve it, which alternately optimizes each block while keeping the other blocks fixed in each iteration. Specifically, the optimization variables are divided into two blocks: $\{\mu_k, \lambda_k\}_{k=1}^K$ and $\{J_k(\mathbf{s})\}_{k=1}^K$. The solutions to the subproblems for each block are provided below.

\subsubsection{Subproblem With Respect to $\{\mu_k, \lambda_k\}_{k=1}^K$}

Given fixed $\{ J_k(\mathbf{s})\}_{k=1}^K$, problem \eqref{problem_no_power_no_fra} is a conventional unconstrained optimization problem with respect to $\{\mu_k, \lambda_k\}_{k=1}^K$, whose optimal solution is given by \cite{shen2018fractional, shen2018fractional2}
\begin{align}
    \label{optimal_mu}
    \mu_k &= \sqrt{1 + \bar{\gamma}_k}, \\
    \label{optimal_lambda}
    \lambda_k &= \frac{\mu_k \int_{\mathcal{S}_{\mathrm{T}}} H_k(\mathbf{s}) J_k(\mathbf{s}) d \mathbf{s}}{\sum_{i=1}^K \left( \left| \int_{\mathcal{S}_{\mathrm{T}}} H_k(\mathbf{s}) J_i(\mathbf{s}) d \mathbf{s} \right|^2 + \frac{\sigma_k^2}{P_{\mathrm{T}}} \int_{\mathcal{S}_{\mathrm{T}}} | J_i(\mathbf{s}) |^2 d \mathbf{s} \right) }.
\end{align} 

\subsubsection{Subproblem With Respect to $\{J_k(\mathbf{s})\}_{k=1}^K$}

Given fixed $\mu_k$ and $\lambda_k$, problem \eqref{problem_no_power_no_fra} can be rewritten as 
\begin{equation}
    \max_{\{J_k(\mathbf{s}) \}_{k=1}^K} \quad \sum_{k=1}^K g(J_k),
\end{equation}
where
\begin{align} \label{eqn_g_function}
    &g \left( J_k \right) = 2 \Re \left\{ A_k \int_{\mathcal{S}_{\mathrm{T}}} H_k (\mathbf{s}) J_k(\mathbf{s}) d \mathbf{s}   \right\} \nonumber \\ & \hspace{0.4cm} - \sum_{i=1}^K \left(B_i \left| \int_{\mathcal{S}_{\mathrm{T}}} H_i(\mathbf{s}) J_k(\mathbf{s}) d \mathbf{s} \right|^2 + C_i \int_{\mathcal{S}_{\mathrm{T}}} | J_k(\mathbf{s}) |^2 d \mathbf{s} \right), 
\end{align}  
where
\begin{equation}
    A_k = \alpha_k \mu_k \lambda_k^*, \quad B_i = \alpha_i |\lambda_i|^2, \quad C_i = \frac{\alpha_i |\lambda_i|^2 \sigma_i^2}{P_{\mathrm{T}}}.
\end{equation}
It can be observed that problem \eqref{eqn_g_function} is a separable optimization with respect to each function $J_k(\mathbf{s})$. Therefore, it can be solved by finding the optimal function $J_k(\mathbf{s})$ that maximizes the functional $g(J_k)$. The CoV is a powerful tool for addressing such functional optimization problems \cite{gelfand2000calculus}. 

Before delving into the details of finding the optimal $J_k$ using the CoV, we first introduce the fundamental lemma of CoV in complex space in the following.
\begin{lemma} \label{fundamental_lemma}
    \normalfont
    \emph{(Fundamental Lemma of CoV in Complex Space)} For every smooth function $U$ defined on an open set $\mathcal{S}$ in the complex space, with the property that
    \begin{equation}
        U(\mathbf{s}) = 0, \forall \mathbf{s} \in \partial \mathcal{S},
    \end{equation}
    with $\partial \mathcal{S}$ being the boundary of $\mathcal{S}$,   
    if a continuous function $V$ on $\mathcal{S}$ satisfies 
    \begin{equation} \label{fundamental_lemma_condition}
        \Re \left\{ \int_{\mathcal{S}} U^* (\mathbf{s}) V (\mathbf{s}) d \mathbf{s} \right\} = 0,
    \end{equation}
    then it must follow that
    \begin{equation}
        V(\mathbf{s}) = 0, \forall \mathbf{s} \in \mathcal{S}.
    \end{equation}
\end{lemma}

\begin{IEEEproof}
    Please refer to Appendix \ref{fundamental_lemma_proof}.
\end{IEEEproof}

Based on \textbf{Lemma \ref{fundamental_lemma}}, we now present the necessary condition for the optimal $J_k$ that maximizes the functional $g(J_k)$ in the following proposition.

\begin{proposition} \label{theorem_optimal_condition}
    \normalfont
    \emph{(Optimal Beamforming Structure)} The function $J_k(\mathbf{s})$ that maximizes the functional $g \left( J_k \right)$ has the following structure:
    \begin{align} \label{optimal_condition}
        J_k(\mathbf{s}) = \bar{A}_k H_k^* (\mathbf{s}) - \sum_{i=1}^K \bar{B}_i H_i^*(\mathbf{s}) \int_{\mathcal{S}_{\mathrm{T}}} H_i (\mathbf{z}) J_k (\mathbf{z}) d \mathbf{z},
    \end{align}
    where 
    \begin{equation}
        \bar{A}_k = \frac{A_k}{\sum_{i=1}^K C_i}, \quad \bar{B}_k = \frac{B_k}{\sum_{i=1}^K C_i}.
    \end{equation}
\end{proposition}

\begin{IEEEproof}
    Please refer to Appendix \ref{theorem_optimal_condition_proof}.
\end{IEEEproof}

The above proposition implies that the optimal $J_k(\mathbf{s})$  is the solution to the equation \eqref{optimal_condition}, which is a Fredholm integral equation of the second kind. To solve it, we rewrite it as 
\begin{align} \label{Fredholm_integral}
    J_k(\mathbf{s}) =\bar{A}_k H_k^*(\mathbf{s}) - \sum_{i=1}^K w_{k,i} \bar{B}_i H_i^*(\mathbf{s}),
\end{align}
where 
\begin{equation} \label{w_defination}
    w_{k,i} \triangleq \int_{\mathcal{S}_{\mathrm{T}}} H_i (\mathbf{z}) J_k (\mathbf{z}) d \mathbf{z}.
\end{equation}
It can be observed from \eqref{Fredholm_integral} that once $w_{k,i}, \forall k, i, $ are calculated, the function $J_k(\mathbf{s})$ can be obtained, since $\bar{A}_k$, $\bar{B}_k$, and $H_k(\mathbf{s}), \forall k$, are known.    Although $w_{k,i}$ is a function of $J_k(\mathbf{s})$, it can be calculated based on the following procedure. Specifically, multiplying both sides of \eqref{Fredholm_integral} by $H_j(\mathbf{s})$ and integrating over $d \mathbf{s}$, we have 
\begin{align} \label{Fredholm_integral_1}
    \int_{\mathcal{S}_{\mathrm{T}}} H_j(\mathbf{s}) J_k(\mathbf{s}) d \mathbf{s} &= \bar{A}_k \int_{\mathcal{S}_{\mathrm{T}}} H_j(\mathbf{s}) H_k^*(\mathbf{s}) d \mathbf{s} \nonumber \\
    &\quad - \sum_{i=1}^K w_{k,i} \bar{B}_i \int_{\mathcal{S}_{\mathrm{T}}} H_j(\mathbf{s}) H_i^*(\mathbf{s}) d \mathbf{s}.
\end{align}   
Note that in the above equation, the left-hand side is essentially $\int_{\mathcal{S}_{\mathrm{T}}} H_j(\mathbf{s}) J_k(\mathbf{s}) d \mathbf{s} =w_{k,j}$. Defining the channel correlation between user $i$ and user $j$ as  
\begin{equation} \label{correlation_ij}
    q_{i,j} \triangleq \int_{\mathcal{S}_{\mathrm{T}}} H_j(\mathbf{s}) H_i^*(\mathbf{s}) d \mathbf{s}, \forall i, j,
\end{equation}
the equation \eqref{Fredholm_integral_1} can be rewritten as 
\begin{equation} \label{eq_31}
    w_{k,j} = \bar{A}_k q_{k,j} - \sum_{i=1}^K \bar{B}_i q_{i,j} w_{k,i}.
\end{equation} 
Then, defining a matrix $\mathbf{W} \in \mathbb{C}^{K \times K}$ whose entry in the $k$-th column and $i$-th row is $\mathbf{W}(k,i) = w_{k,i}$, the equation \eqref{eq_31} can be written in matrix form as 
\begin{equation}
    \mathbf{W} = \mathbf{Q}\mathbf{A} - \mathbf{Q}\mathbf{B} \mathbf{W} \quad \Leftrightarrow \quad (\mathbf{I}_K + \mathbf{Q}\mathbf{B}) \mathbf{W} = \mathbf{Q}\mathbf{A},
\end{equation}
where 
\begin{align}
    \label{update_A}
    &\mathbf{A} = \mathrm{diag}\{ \bar{A}_1,\dots, \bar{A}_K\} \in \mathbb{C}^{K \times K}, \\
    \label{update_B}
    &\mathbf{B} = \mathrm{diag}\{ \bar{B}_1,\dots, \bar{B}_K\} \in \mathbb{R}_+^{K \times K}, \\
    \label{correlation_matrix}
    &\mathbf{Q} = [\mathbf{q}_1,\dots,\mathbf{q}_K] \in \mathbb{S}_+^{K \times K}, \\
    &\mathbf{q}_k = [q_{k,1},\dots,q_{k,K}]^T \in \mathbb{C}^{K \times 1}.
\end{align}
Since the matrix $\mathbf{Q}$ is positive semidefinite and $\mathbf{B}$ is a diagonal matrix with positive entries, the matrix $(\mathbf{I}_K + \mathbf{Q} \mathbf{B})$ is guaranteed to be invertible. Consequently, matrix $\mathbf{W}$ can be calculated as 
\begin{equation} \label{optimal_W}
    \mathbf{W} = (\mathbf{I}_K + \mathbf{Q}\mathbf{B})^{-1} \mathbf{Q}\mathbf{A}.
\end{equation} 
By substituting the entries $w_{k,i}$ of the matrix $\mathbf{W}$ into \eqref{Fredholm_integral}, the optimal function $J_k(\mathbf{s})$ that maximize \eqref{eqn_g_function} can be obtained. The optimality of the obtained solution $J_k(\mathbf{s})$ is given in the following corollary.

\begin{corollary} \label{corollary_optimal}
    \normalfont
    \emph{(Optimality of the Solution)}
    The solution for $J_k(\mathbf{s})$ given by \eqref{Fredholm_integral} and \eqref{optimal_W} is the unique globally optimal solution that maximizes the functional $g(J_k)$.
\end{corollary}

\begin{IEEEproof}
    According to \textbf{Proposition \ref{theorem_optimal_condition}}, the solution in \eqref{Fredholm_integral} is a locally optimal solution that maximizes the functional $g(J_k)$. Given this fact, and the uniqueness of the matrix inversion in \eqref{optimal_W}, the solution in \eqref{Fredholm_integral} must be the unique locally optimal solution, and therefore the globally optimal solution. This completes the proof.
\end{IEEEproof}

\subsubsection{Overall Algorithm}
Based on the solutions developed for $\{\mu_k, \lambda_k\}_{k=1}^K$ and $\{J(\mathbf{s})\}_{k=1}^K$, respectively, problem \eqref{problem_se_max} can be solved by exploiting BCD through the following procedure:
\begin{enumerate}
    \item Initialize $n=0$ and feasible $J_k^{0}(\mathbf{s})$.
    \item Update $\mu_k^{n+1}$ and $\lambda_k^{n+1}$ by \eqref{optimal_mu} and \eqref{optimal_lambda} with $J_k^{n}(\mathbf{s})$.
    \item Update $J_k^{n+1}(\mathbf{s})$ by \eqref{Fredholm_integral} and \eqref{optimal_W} with $\mu_k^{n+1}$ and $\lambda_k^{n+1}$.
    \item Scale the converged $J_k(\mathbf{s})$ according to \eqref{scaled_solution}.
\end{enumerate}
The last step is to guarantee the transmit power limit without influencing the objective value, c.f. \textbf{Lemma \ref{lemma_power}}.
Although each optimization variable is updated using a closed-form solution in the above BCD procedure, the closed-form solutions \eqref{optimal_mu} and \eqref{optimal_lambda} involve integral calculation, which can result in an extremely high computational complexity. To address this issue, we introduce an integral-free but equivalent BCD procedure in the following.

It can be observed that the closed-form solutions \eqref{optimal_mu} and \eqref{optimal_lambda} are all essentially the functions of $w_{k,i}$ defined in \eqref{w_defination}. In particular, according to \eqref{Fredholm_integral}, the total power integral in \eqref{optimal_mu} and \eqref{optimal_lambda} can be reformulated as a function of matrix $\mathbf{W}$, which is given by 
\begin{align}
    \rho(\mathbf{W}) &\triangleq \sum_{k=1}^K \int_{\mathcal{S}_{\mathrm{T}}} | J_k(\mathbf{s}) |^2 d \mathbf{s} \nonumber \\
    &= \sum_{k=1}^K \Bigg( |\bar{A}_k|^2 q_{k,k} 
       -  \sum_{i=1}^K 2\Re \left\{ w^*_{k,i} \bar{A}_k \bar{B}_i^* q_{k,i} \right\}  \nonumber \\
    &\quad + \sum_{i=1}^K \sum_{j=1}^K w_{k,i} w_{k,j}^* \bar{B}_i \bar{B}_j^* q_{i,j} \Bigg) \nonumber \\
    &= \mathrm{tr} \left(\mathbf{A}^H \mathbf{Q} \mathbf{A} - 2 \Re\{\mathbf{A} \mathbf{W}^H \mathbf{B}^H \mathbf{Q}\} + \mathbf{W}^H \mathbf{B}^H \mathbf{Q} \mathbf{B} \mathbf{W}\right).
\end{align}
Therefore, \eqref{optimal_mu} and \eqref{optimal_lambda} can be rewritten as a function of matrix $\mathbf{W}$ as
\setcounter{equation}{38}
\begin{align}
    \label{optimal_mu_closed}
    \mu_k(\mathbf{W}) &= \sqrt{1 + \frac{|w_{k,k}|^2}{\sum_{i=1, i\neq k}^K |w_{i,k}|^2 + \frac{\sigma_k^2}{P_{\mathrm{T}}}\rho(\mathbf{W}) }}, \\
    \label{optimal_lambda_closed}
    \lambda_k(\mathbf{W}) &= \frac{\mu_k(\mathbf{W}) w_{k,k}}{\sum_{i=1}^K |w_{i,k}|^2 + \frac{\sigma_k^2}{P_{\mathrm{T}}} \rho (\mathbf{W}) }.
\end{align}  

With the above new solutions, we only need to iteratively update the $\mathbf{A}$, $\mathbf{B}$, $\mathbf{W}$, $\mu_k$, and $\lambda_k$ during the BCD procedure. Integral calculations are no longer required, except for a few initial integrals computed during the initialization stage. The overall BCD-CoV algorithm for solving problem \eqref{problem_se_max} is summarized in \textbf{Algorithm \ref{alg:BCD_CoV}}.

\begin{algorithm}[tb]
    \caption{BCD-CoV Algorithm for WSR Maximization Problem \eqref{problem_se_max} in CAPA-based Communications}
    \label{alg:BCD_CoV}
    \begin{algorithmic}[1]
        \STATE{calculate the channel correlation matrix $\mathbf{Q}$}
        \STATE{initialize feasible $J_k^{0}(\mathbf{s})$, $\mu_k^{0}$, and $\lambda_k^{0}$, and set $n = 0$ }
        \REPEAT
        \STATE{update $\mathbf{A}^{n}$ and $\mathbf{B}^{n}$ by \eqref{update_A} and \eqref{update_B} with $\mu_k^{n}$ and $\lambda_k^{n}$}
        \STATE{update $\mathbf{W}^{n+1}$ by \eqref{optimal_W} with $\mathbf{A}^{n}$ and $\mathbf{B}^{n}$}
        \STATE{update $\mu_k^{n+1}$ and $\lambda_k^{n+1}$ by \eqref{optimal_mu_closed} and \eqref{optimal_lambda_closed} with $\mathbf{A}^{n}$, $\mathbf{B}^{n}$, and $\mathbf{W}^{n+1}$}
        \STATE{update $n \leftarrow n + 1$}
        \UNTIL{the fractional increase of the objective value of problem \eqref{problem_no_power_no_fra} falls below a threshold}
        \STATE{calculate $J_k(\mathbf{s})$ by \eqref{Fredholm_integral} with the converged $\mathbf{W}$  }
        \STATE{scale $J_k(\mathbf{s})$ according to \eqref{scaled_solution}}
    \end{algorithmic}
\end{algorithm}

\subsection{Initialization, Convergence, and Complexity}
\begin{figure}[t!]
    \centering
    \includegraphics[width=0.45\textwidth]{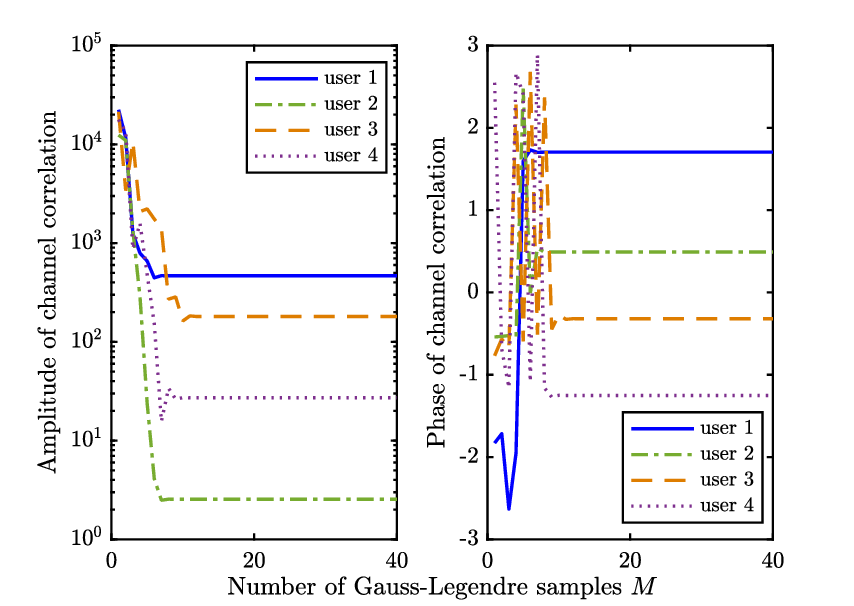}
    \caption{Convergence of Gauss-Legendre quadrature for calculating the channel correlation between a reference user located at $\mathbf{r}_0 = [1,1,30]^T$ and four other users, with $L_x = L_y = \sqrt{0.5}$ m and $f = 7$ GHz.} \label{fig_GL}
    \vspace{0.5cm}
\end{figure}

During the initialization phase of the proposed BCD-CoV algorithm, the channel correlation matrix $\mathbf{Q}$, with entries defined in \eqref{correlation_ij}, needs to be computed. This requires calculating $\frac{K(K+1)}{2}$ integrals, given that $\mathbf{Q} = \mathbf{Q}^H$. These integrals can be computed using Gauss-Legendre quadrature, which takes the form \cite{olver2010nist}:
\begin{equation}
    \int_{a}^{b} \psi(x) dx \approx \frac{b-a}{2} \sum_{m=1}^M \omega_m \psi\left(\frac{b-a}{2}\theta_m + \frac{a+b}{2}\right),
\end{equation} 
where $M$ is the number of sample points, $\omega_m$ are the quadrature weights, and $\theta_m$ are the roots of the $M$-th Legendre polynomial. A larger value of $M$ results in higher approximation accuracy. In the following, we provide an example of computing the entries of $\mathbf{Q}$, assuming the transmit CAPA is deployed in the $x$-$y$ plane and centered at the origin, without loss of generality. Let $L_x$ and $L_y$ denote the length of the CAPA along the $x$- and $y$-axes, respectively. The entry $q_{i,j}$ can be computed by 
\begin{align}
    q_{i,j} &= \int_{\mathcal{S}_{\mathrm{T}}} H_j(\mathbf{s}) H_i^*(\mathbf{s}) d \mathbf{s} \nonumber \\
    &= \int_{-\frac{L_y}{2}}^{\frac{L_y}{2}} \int_{-\frac{L_x}{2}}^{\frac{L_x}{2}} H_j(s_x, s_y) H_i^*(s_x, s_y) d s_x d s_y \nonumber \\
    &\approx \frac{L_x L_y}{4} \sum_{m_y=1}^M \sum_{m_x=1}^M \omega_{m_x} \omega_{m_y} H_j\left(\frac{\theta_{m_x} L_x}{2}, \frac{\theta_{m_y} L_y}{2}\right) \nonumber \\
    & \hspace{3.4cm} \times H_i^*\left(\frac{\theta_{m_x} L_x}{2}, \frac{\theta_{m_y} L_y}{2}\right),
\end{align} 
where the last step is obtained using the Gauss-Legendre quadrature. A numerical example of using Gauss-Legendre quadrature to calculate the channel correlations is shown in Fig. \ref{fig_GL}. It can be observed that using only $M = 10$ sample points is sufficient for accurately calculating the channel correlations.

Additionally, to initialize the source current pattern $J_k(\mathbf{s})$, the matched filtering method can be used, which simply aligns the source current pattern with the channel for each user. Therefore, given that the power constraint has been removed in problem \eqref{problem_no_power_no_fra}, the source current pattern is initialized as
\begin{equation}
    J_k^{0}(\mathbf{s}) = H_k^*(\mathbf{s}).
\end{equation}
Then, by substituting the above solution into \eqref{optimal_mu_closed} and \eqref{optimal_lambda_closed}, the initial auxiliary variables $\mu_k$ and $\lambda_k$ can be obtained as 
\begin{equation}
    \mu_k^0 = \mu_k(\mathbf{Q}), \quad \lambda_k^0 = \lambda_k(\mathbf{Q}).
\end{equation}  
In addition to the matched filtering method, the ZF method, which will be discussed in detail in Section \ref{sec:ZF}, is also an effective method for determining the initial point.

The convergence of the proposed BCD-CoV algorithm is analyzed as follows. Let $f_{\mathrm{obj}} (\mu_k, \lambda_k, J_k)$ denote the objective function of problem \eqref{problem_no_power_no_fra}. For two consecutive iterations of the proposed BCD-CoV algorithm, we have 
\begin{align}
    f_{\mathrm{obj}} (\mu_k^{n+1}, \lambda_k^{n+1}, J_k^{n+1}) & \overset{(a)}{\ge} f_{\mathrm{obj}} (\mu_k^{n+1}, \lambda_k^{n+1}, J_k^{n}) \nonumber \\
    &\overset{(b)}{\ge} f_{\mathrm{obj}} (\mu_k^{n}, \lambda_k^{n}, J_k^{n}),
\end{align}
where step $(a)$ follows from the global optimality of the obtained $J_k(\mathbf{s})$ in each iteration, as proved in \textbf{Corollary \ref{corollary_optimal}}, and step $(b)$ follows a similar process to that in \cite[Appendix A]{shen2018fractional2}. Since the objective function is bounded from above, the strict convergence of the proposed BCD-CoV algorithm is guaranteed.

The complexity of the proposed BCD-CoV algorithm primarily arises from two components. The first is the computation of the channel correlation matrix $\mathbf{Q}$ during the initialization phase. When using $M$-point Gauss-Legendre quadrature, the computational complexity of this step is $O(M^2 K^2)$. The second component is the matrix inversion required for calculating the matrix $\mathbf{W}$ during each iteration, which has a worst-case computational complexity of $O(I_o K^3)$ with $I_o$ being the number of iterations. 
\section{Corr-ZF Approach} \label{sec:ZF}

In this section, we propose a Corr-ZF approach to further reduce the computational complexity of CAPA beamforming. For conventional MIMO systems with spatially discrete antenna arrays, the ZF beamformer has been shown to asymptotically approach the optimal capacity in multi-user systems, while providing closed-form and tractable beamformer expressions \cite{4599181, 6832894}. By imposing an additional constraint on problem \eqref{problem_se_max} to completely eliminate inter-user interference, the ZF problem for maximizing the WSR is formulated as follows:
\begin{subequations} \label{problem_ZF}
    \begin{align}
        \max_{\{ J_k(\mathbf{s}) \}_{k=1}^K} \quad &\sum_{k=1}^K \alpha_k \log_2 \left( 1 + \gamma_k \right) \\
        \mathrm{s.t.} \quad & \sum_{k=1}^K \int_{\mathcal{S}_{\mathrm{T}}} | J_k(\mathbf{s}) |^2 d \mathbf{s} = P_{\mathrm{T}}, \\
        \label{ZF_constraint}
        & \int_{\mathcal{S}_{\mathrm{T}}} {H}_i(\mathbf{s}) J_k (\mathbf{s}) d \mathbf{s} = 0, \forall i \neq k,
    \end{align}
\end{subequations}
where \eqref{ZF_constraint} is the ZF constraint for CAPAs.

Typically, the conventional ZF beamformer is computed using the pseudoinverse of a matrix composed of the channel vectors for all users. However, this approach does not apply to CAPA systems, where the dimension of the channel vectors becomes infinite. To address this challenge, we propose a closed-form Corr-ZF solution for the source current patterns, i.e., the infinite-dimensional beamformer, based on the channel correlation matrix instead, which eliminates the requirement for calculating the pseudoinverse of infinite-dimension matrices. The proposed Corr-ZF solution is given in the following proposition.

\begin{proposition} \label{proposition_ZF}
    \normalfont
    \emph{(ZF Beamforming)}
    The function $\{J_k(\mathbf{s})\}_{k=1}^K$ that achieves the ZF constraint \eqref{ZF_constraint} can be expressed as 
    \begin{align} \label{ZF_beamformer}
        J_k(\mathbf{s}) &= \sqrt{\rho_k} J_k^{\mathrm{ZF}}(\mathbf{s}), \\ J_k^{\mathrm{ZF}}(\mathbf{s}) &= \sum_{j=1}^K u_{k,j} H_j^*(\mathbf{s}),
    \end{align}
    where $\rho_k$ is a power scaling factor, $u_{k,j}$ is the entry of $\mathbf{Q}^{-1}$ in the $k$-th column and $j$-th row, and $\mathbf{Q}$ is the channel correlation matrix defined in \eqref{correlation_matrix}. 
\end{proposition}

\begin{IEEEproof}
    Substituting \eqref{ZF_beamformer} into \eqref{ZF_constraint} yields 
    \begin{align}
        &\int_{\mathcal{S}_{\mathrm{T}}} H_i(\mathbf{s}) J_k(\mathbf{s}) d \mathbf{s} \nonumber \\
        &= \sqrt{\rho_k} \sum_{j=1}^K u_{k,j} \int_{\mathcal{S}_{\mathrm{T}}} H_i(\mathbf{s}) H_j^*(\mathbf{s}) d \mathbf{s} = \sqrt{\rho_k} \bar{\mathbf{q}}_i^T \mathbf{u}_k,
    \end{align}
    where $\bar{\mathbf{q}}_i = [q_{1,i},\dots,q_{K,i}]^T$ is the $i$-th column of $\mathbf{Q}^T$ and $\mathbf{u}_k = [u_{k,1},\dots,u_{k,K}]^T$ is the $k$-th column of $\mathbf{Q}^{-1}$. Since $\mathbf{Q} \mathbf{Q}^{-1} = \mathbf{I}_K$, we must have
    \begin{equation} \label{ZF_ortho}
        \bar{\mathbf{q}}_i^T \mathbf{u}_k = \begin{cases}
            1, & i = k, \\
            0, & i \neq k.
        \end{cases}
    \end{equation}    
    This completes the proof.
\end{IEEEproof}

\begin{table*}
    \centering
    \caption{Comparison of Proposed Approaches and Fourier-based Approaches.}
    \label{table_compare}
    \begin{tabular}{c|c|c|c|c}
      \toprule
      {Approaches} & {\makecell[c]{Approximations\\Applied?}} & {\makecell[c]{Initialization\\Complexity}} & {\makecell[c]{Optimization\\Complexity}} & \makecell[c]{Beamforming\\ Structure}   \\
        \midrule
      Proposed CoV-based approach & No & $O(M^2 K^2)$ & $O(I_o K^3)$ & Optimal  \\
      Proposed Corr-ZF approach & No & $O(M^2 K^2)$ & $O(K^3) + O(K)$ & Asymptotically Optimal \\
      Fourier-based approach & Yes & $O(M^2 K N_{\mathrm{F}})$ & $O(I_o N_{\mathrm{F}}^3)$ & Non-Optimal  \\
      Fourier-based ZF approach & Yes & $O(M^2 K N_{\mathrm{F}})$ & $O(K^2 N_{\mathrm{F}}) + O(K)$ & Non-Optimal\\
      \bottomrule
      \multicolumn{5}{l}{Some typical values: $M \propto 10^1, K \propto 10^1,$ and $N_{\mathrm{F}} \propto  10^2 \sim 10^4$. }
    \end{tabular}
\end{table*}

By further taking into account the power constraint, the power scaling factor in \eqref{ZF_beamformer} can be designed as 
\begin{equation} \label{ZF_power}
    \rho_k = \frac{P_k}{\int_{\mathcal{S}_{\mathrm{T}}} | J_k^{\mathrm{ZF}}(\mathbf{s}) |^2 d \mathbf{s}} = \frac{P_k}{ \mathbf{u}_k^H \mathbf{Q} \mathbf{u}_k } = \frac{P_k}{u_{k,k}},
\end{equation}
where the last step is obtained from \eqref{ZF_ortho} and $P_k$ is the power allocating to user $k$, satisfying $\sum_{k=1}^K P_k = P_{\mathrm{T}}$. Given the ZF solution given in \eqref{ZF_beamformer} and \eqref{ZF_power}, the SINR for user $k$ becomes
\begin{align}
    \gamma_k^{\mathrm{ZF}} = &\frac{\rho_k \left| \int_{\mathcal{S}_{\mathrm{T}}} H_k(\mathbf{s}) J_k^{\mathrm{ZF}}(\mathbf{s}) d \mathbf{s}  \right|^2 }{\rho_k \sum_{i=1,i\neq k}^K  \left| \int_{\mathcal{S}_{\mathrm{T}}} H_i(\mathbf{s}) J_k^{\mathrm{ZF}}(\mathbf{s}) d \mathbf{s}  \right|^2 + \sigma_k^2} \nonumber \\
    = &\frac{P_k | \bar{\mathbf{q}}_k^T\mathbf{u}_k|^2}{P_k \sum_{i=1, i \neq k}^K |\bar{\mathbf{q}}_i^T\mathbf{u}_k|^2 + u_{k,k} \sigma_k^2 } = \frac{P_k}{u_{k,k}\sigma_k^2},
\end{align}
where the last step is obtained from \eqref{ZF_ortho}.
Therefore, problem \eqref{problem_ZF} reduces to the following power allocation problem:
\begin{subequations} \label{problem_ZF_reduced}
    \begin{align}
        \max_{\{ P_k \}_{k=1}^K } \quad &\sum_{k=1}^K \alpha_k \log \left( 1 + \frac{P_k}{u_{k,k} \sigma_k^2} \right) \\
        \mathrm{s.t.} \quad & \sum_{k=1}^K P_k = P_{\mathrm{T}}.
    \end{align}
\end{subequations}

\begin{proposition} \label{proposition_WF}
    \normalfont
    \emph{(Power Allocation)}
    The optimal solution to problem \eqref{problem_ZF_reduced} is
    \begin{equation}
        P_k = \left( \nu \alpha_k - u_{k,k} \sigma_k^2 \right)^{+},
    \end{equation}
    where
    \begin{equation}
        \nu = \frac{1}{\sum_{k=1}^M \alpha_k} \left( P_{\mathrm{T}} + \sum_{k=1}^M u_{k,k} \sigma_k^2 \right),
    \end{equation}
    and $M$ is the number of non-zero $P_k$.
\end{proposition}

\begin{IEEEproof}
    The above solution is the classical water-filling solution that can be obtained based on the Karush-Kuhn-Tucker conditions. We thus omit the details here.
\end{IEEEproof}

The water-filling solution given in \textbf{Proposition \ref{proposition_WF}} can be calculated using the method in \cite[Corollary 1]{1381759}, which has a worst-case complexity of $O(K)$. Additionally, the complexity of the proposed Corr-ZF approach also arises from calculating the inverse of the matrix $\mathbf{Q}$, c.f. \textbf{Proposition \ref{proposition_ZF}}, which has a worst-case complexity of $O(K^3)$. 


\section{Comparison with Fourier-based Approach} \label{sec:Fourier_method}
In this section, we compare the proposed approaches with the state-of-the-art Fourier-based approach. The core idea of this approach is to approximate continuous functions using a finite number of Fourier series terms. Specifically, according to \cite[Lemma 2]{zhang2023pattern}, the source current patterns can be expressed as a Fourier series:
\begin{equation} \label{Fourier_J}
    J_k(\mathbf{s}) = \sum_{\mathbf{n}=-\infty}^{\infty} v_{k,\mathbf{n}} \Phi_{\mathbf{n}}(\mathbf{s}),
\end{equation}
where $\mathbf{n} = [n_x, n_y, n_z]^T$, and the sum notation is defined as $\sum_{\mathbf{n}=-\infty}^{\infty} \triangleq \sum_{n_x = -\infty}^{\infty} \sum_{n_y = -\infty}^{\infty} \sum_{n_z = -\infty}^{\infty}$. The Fourier coefficients and the orthonormal Fourier basis functions are given by
\begin{align}
    v_{k, \mathbf{n}} &= \frac{1}{\sqrt{A_{\mathrm{T}}}} \int_{\mathcal{S}_{\mathrm{T}}} J_k(\mathbf{s}) \Phi_{\mathbf{n}}^*(\mathbf{s}) d \mathbf{s}, \\
    \Phi_{\mathbf{n}}(\mathbf{s}) &= \frac{1}{\sqrt{A_{\mathrm{T}}}} e^{j 2 \pi \left( \frac{n_x}{L_x}\left( s_x - \frac{L_x}{2} \right) + \frac{n_y}{L_y}\left( s_y - \frac{L_y}{2} \right) + \frac{n_z}{L_z}\left( s_z - \frac{L_z}{2} \right) \right) },
\end{align}
where $\mathbf{s} = [s_x, s_y, s_z]^T$, and $L_x$, $L_y$, and $L_z$ are the projection lengths of the aperture $\mathcal{S}_{\mathrm{T}}$ onto the $x$-, $y$-, and $z$-axes, respectively. To address the infinite sum in the above Fourier series, the following finite-sum approximations are typically used \cite[Proposition 1]{zhang2023pattern}:
\begin{align}
    J_k(\mathbf{s}) \approx \sum_{\mathbf{n}=-\mathbf{N}}^{\mathbf{N}} v_{k,\mathbf{n}} \Phi_{\mathbf{n}}(\mathbf{s}),
\end{align} 
where the sum notation is defined as $\sum_{\mathbf{n}=-\mathbf{N}}^{\mathbf{N}} \triangleq \sum_{n_x = -N_x}^{N_x} \sum_{n_y = -N_y}^{N_y} \sum_{n_z = -N_z}^{N_z}$ with 
\begin{equation}
    N_x = \left\lceil \frac{L_x}{\lambda} \right\rceil, \quad N_y = \left\lceil \frac{L_y}{\lambda} \right\rceil, \quad N_z = \left\lceil \frac{L_z}{\lambda} \right\rceil. 
\end{equation}
Then, we have the following approximations:
\begin{align}
    \int_{\mathcal{S}_{\mathrm{T}}} |J_k(\mathbf{s})|^2 d \mathbf{s} &\approx \sum_{\mathbf{n}=-\mathbf{N}}^{\mathbf{N}} |v_{k,\mathbf{n}}|^2 = \| \mathbf{v}_k \|^2 \\ 
    \int_{\mathcal{S}_{\mathrm{T}}} H_i(\mathbf{s}) J_k(\mathbf{s}) d \mathbf{s} &\approx \sum_{\mathbf{n}=-\mathbf{N}}^{\mathbf{N}} g_{i, \mathbf{n}} v_{k,\mathbf{n}} = \mathbf{g}_i^T \mathbf{v}_k,  
\end{align}
where $g_{i, \mathbf{n}}$ is the Fourier transform of $H_i(\mathbf{s})$, given by 
\begin{equation} \label{Fourier_channel}
    g_{i, \mathbf{n}} = \int_{\mathcal{S}_{\mathrm{T}}} H_i(\mathbf{s}) \Phi_{\mathbf{n}}(\mathbf{s}) d \mathbf{s},
\end{equation} 
and $\mathbf{v}_k \in \mathbb{C}^{N_{\mathrm{F}} \times 1}$ and $\mathbf{g}_i \in \mathbb{C}^{N_{\mathrm{F}} \times 1}$, with $N_{\mathrm{F}} = (2N_x+1)(2N_y+1)(2N_z+1)$,  are vectors collecting all $v_{k,\mathbf{n}}$ and $g_{i, \mathbf{n}}, \forall \mathbf{n} \in [-\mathbf{N}, \mathbf{N}]$, respectively.
Based on the above approximations, the WSR maximization problem in \eqref{problem_se_max} can be approximated as 
\begin{subequations} \label{problem_se_max_approx}
    \begin{align}
        \max_{\{ \mathbf{v}_k \}_{k=1}^K} \quad &\sum_{k=1}^K \alpha_k \log_2 \left( 1 + \frac{\left| \mathbf{g}_k^T \mathbf{v}_k \right|^2 }{\sum_{i=1,i\neq k}^K \left| \mathbf{g}_k^T \mathbf{v}_i \right|^2 + \sigma_k^2} \right) \\
        \label{Fourier_power}
        \mathrm{s.t.} \quad & \sum_{k=1}^K \|\mathbf{v}_k\|^2 \le P_{\mathrm{T}}.
    \end{align}
\end{subequations}
The above problem is the classical WSR maximization for spatially discrete antenna arrays, which can be solved using traditional methods, such as fractional programming (FP) techniques \cite{shen2018fractional, shen2018fractional2} or the classical weighted minimum mean-squared error (WMMSE) method \cite{4712693}. 

\begin{table*}
    \centering
    \caption{Comparison of Average CPU Time for Different Approaches.}
    \label{table_CPU_time}
    \begin{tabular}{c|cc|cc|cc|cc}
      \toprule
      \multirow{2}{*}{Frequency} &
        \multicolumn{2}{c|}{$A_{\mathrm{T}} = 0.2$ m$^2$} &
        \multicolumn{2}{c|}{$A_{\mathrm{T}} = 0.3$ m$^2$} &
        \multicolumn{2}{c|}{$A_{\mathrm{T}} = 0.4$ m$^2$} &
        \multicolumn{2}{c}{$A_{\mathrm{T}} = 0.5$ m$^2$} \\
        & CoV & Fourier & CoV & Fourier & CoV & Fourier & CoV & Fourier \\
        \midrule
      2.4 GHz & 0.147 s & 0.267 s & 0.149 s & 0.323 s & 0.148 s & 0.543 s & 0.147 s & 0.545 s \\
      7.8 GHz & 0.148 s & 1.971 s & 0.148 s & 3.074 s & 0.148 s & 3.985 s & 0.148 s & 5.028 s \\
      15 GHz & 0.149 s & 7.526 s & 0.149 s & 11.944 s & 0.148 s & 16.808 s & 0.147 s & 23.415 s\\
      \bottomrule
    \end{tabular}
\end{table*}

The computational complexity of the Fourier-based approach is analyzed as follows. First, before solving problem \eqref{problem_se_max_approx}, all Fourier transforms of the channel need to be computed as given in \eqref{Fourier_channel}, which requires calculating $K N_{\mathrm{F}}$ integrals. Gauss-Legendre quadrature can be employed to compute each integral. When using $M$-point Gauss-Legendre quadrature, the computational complexity of this step is $O(M^2 K N_{\mathrm{F}})$. Next, the computational complexity for solving problem \eqref{problem_se_max_approx} using the FP or WMMSE method is $O(I_o N_{\mathrm{F}}^3)$. Compared to the proposed BCD-CoV algorithm, whose complexity is primarily determined by $K$, the complexity of the Fourier-based approach is mainly driven by $N_{\mathrm{F}}$. According to the definition of $N_{\mathrm{F}}$ given below \eqref{Fourier_channel}, its value increases with both the CAPA aperture size and the carrier frequency. Notably, the value of $N_{\mathrm{F}}$ can become extremely large under practical system settings. For example, consider a CAPA with $L_x = L_y = 0.5$ m and $L_z = 0$. In this case, we have $N_{\mathrm{F}} = 81$ at $2.4$ GHz, $N_{\mathrm{F}} = 729$ at $7.8$ GHz, and $N_{\mathrm{F}} = 2601$ at $15$ GHz. Therefore, the complexity of the Fourier-based approach can be extremely high when a large aperture array or high-frequency band is used. In addition to reducing computational complexity, the proposed CoV-based approach directly optimizes the continuous source current patterns rather than their approximations, thereby offering improved performance.

To reduce the computational complexity, problem \eqref{problem_se_max_approx} can also be addressed under the following ZF constraint:
\begin{equation}
    |\mathbf{g}_i^T \mathbf{v}_k| = 0, \forall i \neq k,
\end{equation}
with solutions provided in \cite{4599181, 6832894}. These conventional ZF solutions require calculating the pseudoinverse of an $N_{\mathrm{F}} \times K$ channel matrix, with entries given in \eqref{Fourier_channel}, resulting in a computational complexity of $O(K^2 N_{\mathrm{F}})$. This is followed by a water-filling process for power allocation, which has a complexity of $O(K)$. In Table \ref{table_compare}, we summarize the comparison of proposed approaches and Fourier-based approaches.

\section{Numerical Results} \label{sec:results}

In this section, numerical results obtained through Monte Carlo simulations are provided to verify the effectiveness of the proposed CoV-based and Corr-ZF approaches for CAPA beamforming. The following simulation setup, similar to that in \cite{zhang2023pattern}, is used throughout our simulations unless stated otherwise. It is assumed that the transmit CAPA is deployed within the $x$-$y$ plane and centered at the origin of the coordinate system, where 
\begin{equation}
  \mathcal{S}_{\mathrm{T}} = \left\{ [s_x, s_y, 0]^T \Big| |s_x| \le \frac{L_x}{2}, |s_y| \le \frac{L_y}{2} \right\},
\end{equation}
with $L_x = L_y = \sqrt{A_{\mathrm{T}}}$ and $A_{\mathrm{T}} = 0.25 \text{ m}^2$. There are $K = 8$ communication users randomly located within the following region 
\begin{equation}
  \mathcal{U} = \left\{  [r_x, r_y, r_z]^T \Bigg| \begin{matrix}
    &|r_x| \le U_x, |r_y| \le U_y,\\
    &U_{z, \min} \le r_z \le U_{z, \max}
  \end{matrix} \right\},
\end{equation} 
with $U_x = U_y = 5$ m, $U_{z, \min} = 15$ m, and $U_{z, \max} = 30$ m. The polarization direction of all users are set to $\hat{\mathbf{u}}_x = \hat{\mathbf{u}}_y = [0,1,0]^T, \forall k$. The signal frequency and the intrinsic impedance are set to $f = 2.4$ GHz and $\eta = 120 \pi$ $\Omega$, respectively. The transmit power and the noise power are to $P_{\mathrm{T}} = 100$ mA$^2$ and $\sigma_k^2 = 5.6 \times 10^{-3} \text{ V}^2/\text{m}^2$, respectively. The weight of each user is set to $\alpha_k = 1/K, \forall k$.  The number of samples of the Gauss-Legendre quadrature for calculating all integrals in the simulation is set to $M = 20$. All following results are obtained by
averaging over $200$ random channel realizations unless otherwise specified. 

For performance comparison, we mainly consider the following benchmark schemes.
\begin{itemize}
  \item \textbf{Fourier-based approach}: In this benchmark, the Fourier-based approach described in Section \ref{sec:Fourier_method} is adopted, where the continuous source current pattern is approximated using finite Fourier series. 
  \item \textbf{Conventional MIMO}: In this benchmark, the conventional spatially discrete antenna array is exploited, where the continuous surface $\mathcal{S}_{\mathrm{T}}$ is occupied with discrete antennas with $A_d = \frac{\lambda^2}{4\pi}$ effective aperture area of each antenna and $d = \frac{\lambda}{2}$ antenna spacing. The location of the $(n_x, n_y)$-th antenna is given by
  \begin{equation}
    \bar{\mathbf{s}}_{n_x, n_y} = \left[ (n_x-1)d - \frac{L_x}{2}, (n_y-1)d - \frac{L_y}{2}, 0 \right]^T.
  \end{equation} 
  Therefore, there are totally $N_d = \lceil \frac{L_x}{d} \rceil \times \lceil \frac{L_y}{d} \rceil$ antennas. Let $\mathcal{S}_{n_x, n_y}$ denote the surface of the $(n_x, n_y)$-th antenna, where $|\mathcal{S}_{n_x, n_y}| = A_d$. The channel between this antenna and user $k$ can be calculated as~\cite{9906802, zhang2023pattern}
  \begin{align}
    h_{k, n_x, n_y} = &\frac{1}{\sqrt{A_d}} \int_{\mathcal{S}_{n_x, n_y}} \hspace{-0.6cm} H_k(\mathbf{s}) d \mathbf{s} 
    \approx \sqrt{A_d} H_k (\bar{\mathbf{s}}_{n_x, n_y}) .
  \end{align}
  Given the above discrete channels, the conventional MIMO beamforming optimization problem for maximizing WSR can be solved using traditional methods, such as FP and WMMSE methods. The ZF method can also be exploited for conventional MIMO to reduce the beamforming complexity \cite{4599181, 6832894}.

\end{itemize}

\begin{figure}[t!]
  \centering
  \includegraphics[width=0.45\textwidth]{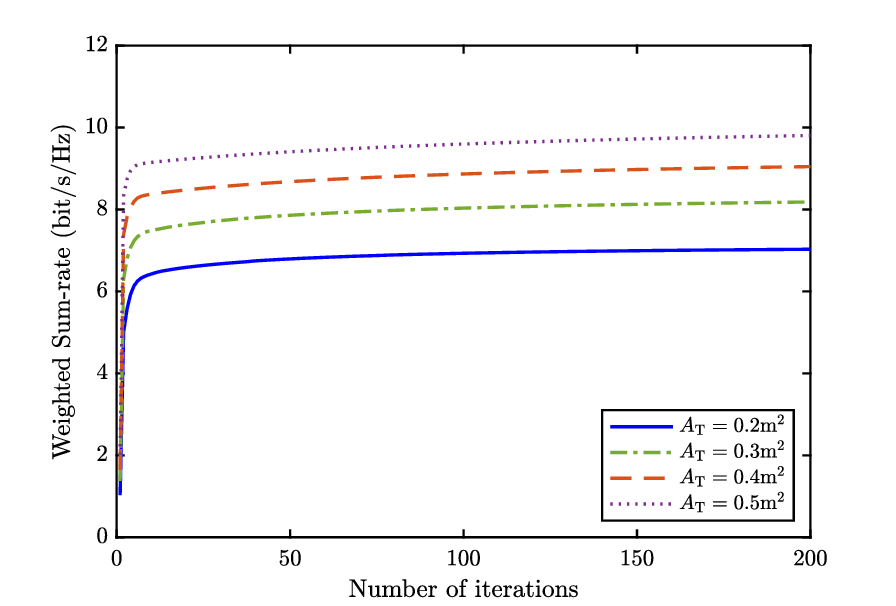}
  \caption{Convergence of the proposed BCD-CoV algorithm.}
  \label{fig_convergence}
\end{figure}

\subsection{Convergence and Complexity of CoV Approach}
Fig. \ref{fig_convergence} illustrates the average convergence behavior of the proposed BCD-CoV algorithm for different aperture sizes. It can be observed that the BCD-CoV algorithm consistently converges to a stable value after a few iterations. Table \ref{table_CPU_time} further compares the CPU time consumed by the proposed CoV-based approach with the state-of-the-art Fourier-based approach, using MATLAB R2024a on an Apple M3 silicon chip. From this table, we observe the following: for the Fourier-based approach, CPU time increases dramatically with the aperture size and frequency. For instance, as the aperture size and frequency increase from $0.2 \text{ m}^2$ and $2.4$ GHz to $0.5 \text{ m}^2$ and $15$ GHz, the CPU time consumed by the Fourier-based approach rises by almost $100$ times, from $0.267$ seconds to $23.415$ seconds. This is because the required number of reserved Fourier series items, i.e., $N_{\mathrm{F}}$, is extremely high when either the aperture size is large or the frequency is high. In contrast, the CPU time consumed by the proposed CoV-based approach remains consistently low, around $0.148$ seconds, as its complexity is independent of both aperture size and frequency. These results validate the efficiency of the CoV-based approach in directly optimizing the continuous source current pattern.

\begin{figure}[t!]
  \centering
  \includegraphics[width=0.45\textwidth]{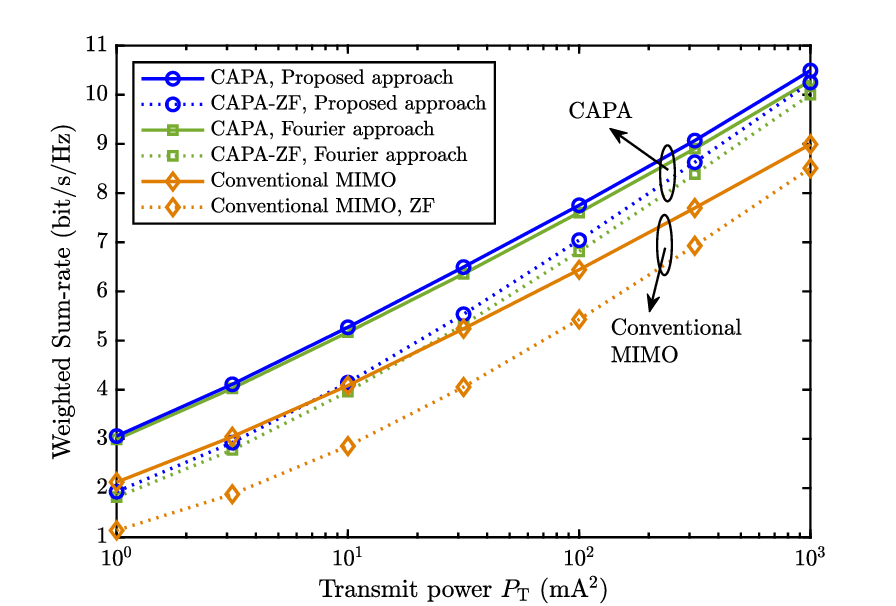}
  \caption{WSR versus transmit power.}
  \label{fig_transmit_power}
\end{figure} 

\begin{figure}[t!]
  \centering
  \includegraphics[width=0.45\textwidth]{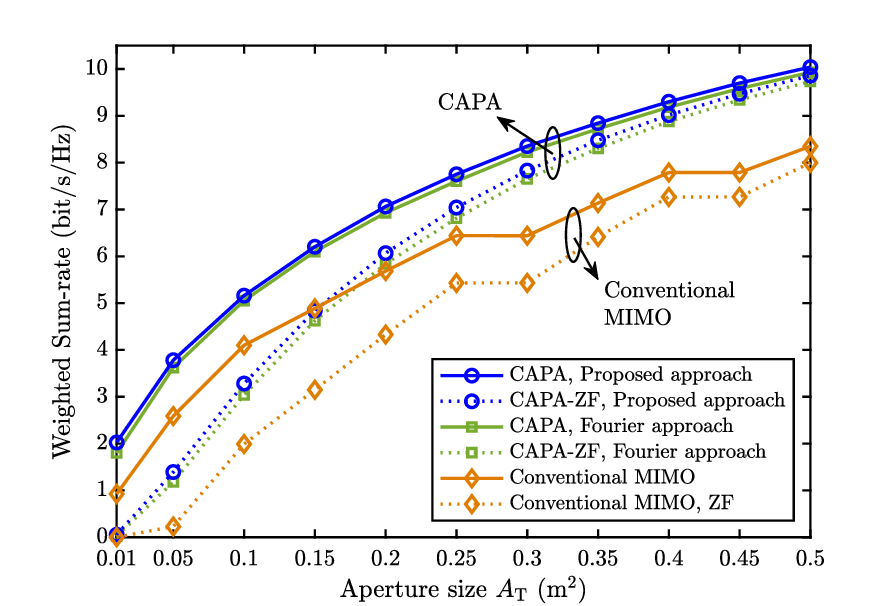}
  \caption{WSR versus aperture size.}
  \label{fig_aperture_size}
\end{figure}

\subsection{WSR Versus Transmit Power}

Fig. \ref{fig_transmit_power} illustrates the impact of transmit power $P_{\mathrm{T}}$ on the WSR performance. It can be observed that the WSR achieved by all approaches increases with the transmit power, while the proposed approaches always outperform the benchmarks. For example, when $P_{\mathrm{T}} = 10^3 \text{ mA}^2$, using CAPA with the proposed approach increases the WSR by $17\%$, from $9$ to $10.5$ bit/s/Hz, compared to conventional MIMO without ZF, and by $20\%$, from $8.5$ to $10.2$ bit/s/Hz, with ZF. Although the Fourier-based approaches can achieve WSR performance comparable to the proposed approaches, both with and without ZF, this performance comes at the cost of significantly higher computational complexity, as shown in Table \ref{table_CPU_time}. Additionally, the WSR performance with ZF becomes closer to that without ZF as the transmit power increases. This is expected, as inter-user interference dominates over noise in the high-power regime, making the elimination of interference more critical. Conversely, in the low-power regime, noise dominates over inter-user interference, rendering ZF less effective.
 
\begin{figure}[t!]
  \centering
  \includegraphics[width=0.45\textwidth]{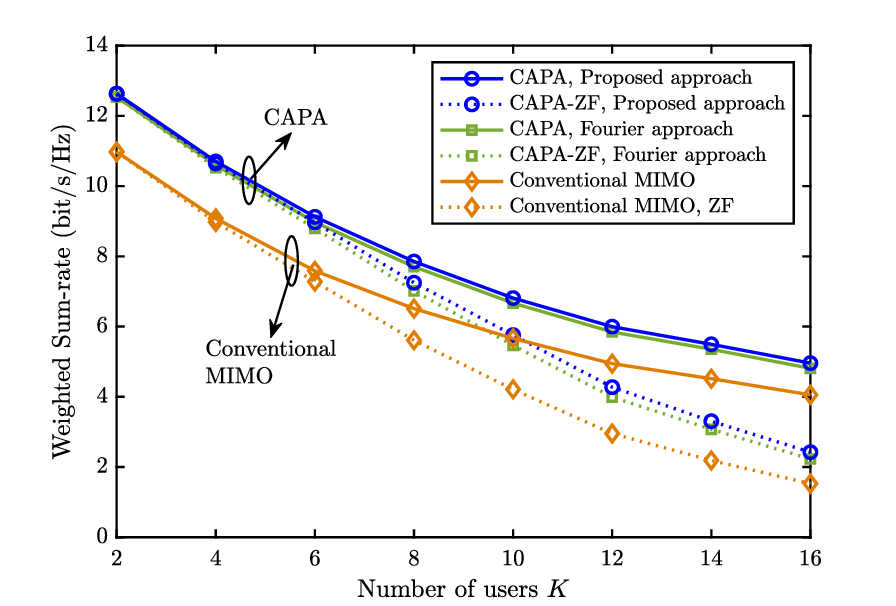}
  \caption{WSR versus the number of users.}
  \label{fig_user}
\end{figure} 

\begin{figure}[t!]
  \centering
  \includegraphics[width=0.45\textwidth]{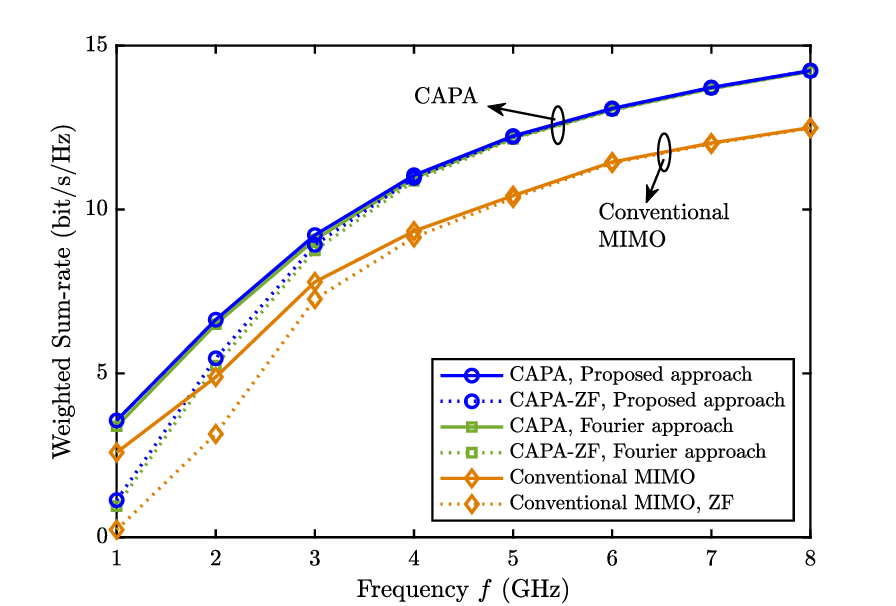}
  \caption{WSR versus carrier frequency.}
  \label{fig_frequency}
\end{figure} 

\subsection{WSR Versus Aperture Size}

Fig. \ref{fig_aperture_size} studies the impact of aperture size $A_{\mathrm{T}}$ on the WSR performance. It can be observed that the WSR achieved by all approaches increases with the aperture size. This is because increasing the aperture size improves the DoFs for both enhancing signal strength and mitigating inter-user interference. Moreover, the proposed approaches consistently achieve the best WSR performance across all aperture sizes. For instance, compared to conventional MIMO, CAPA with the proposed approach increases the WSR by $21\%$, from $6.4$ to $7.8$ bit/s/Hz without ZF, and by $30\%$, from $5.4$ to $7.0$ bit/s/Hz with ZF, when $A_{\mathrm{T}} = 0.25 \text{ m}^2$. The improvement of WSR achieved by CAPA over conventional MIMO becomes more significant as the aperture size grows, indicating the strong ability of continuous transmit pattern to fully exploit the DoFs provided by the larger aperture. Similarly, the Fourier-based approach exhibits close performance to the proposed approach at a cost of high computational complexity. Furthermore, the performance gap between the WSRs achieved with and without ZF gradually diminishes as the aperture size increases. This is because a larger aperture size makes the channels between different users closer to orthogonal, allowing ZF to better align the transmit patterns with the channels of different users.

\begin{figure}[t!]
  \centering
  \begin{subfigure}[t]{0.5\textwidth}
  \includegraphics[width=1\textwidth]{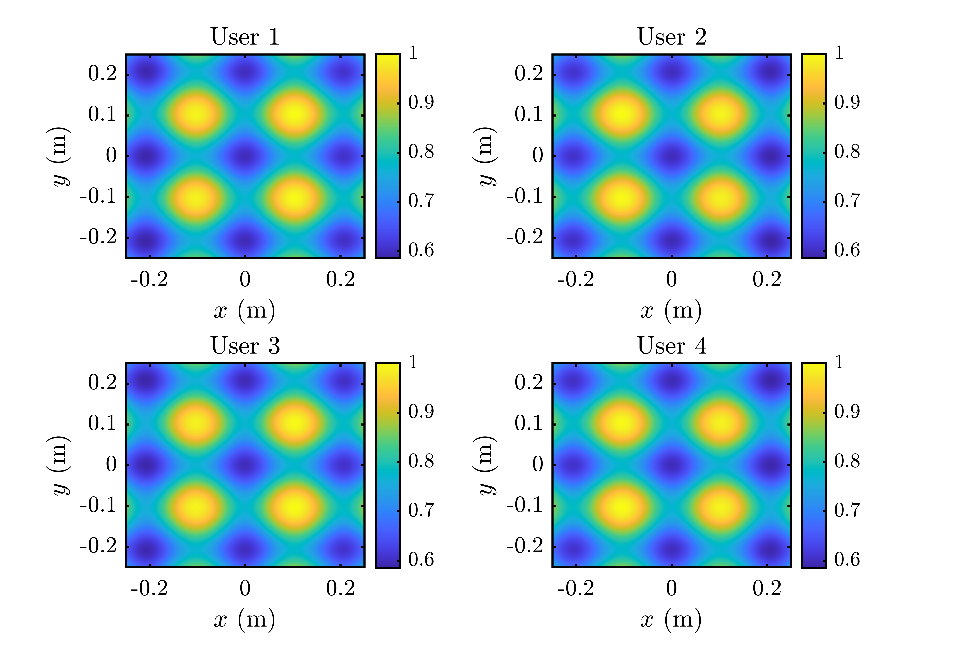}
  \caption{Amplitudes of the source current patterns.}
  \end{subfigure}
  \begin{subfigure}[t]{0.5\textwidth}
  \includegraphics[width=1\textwidth]{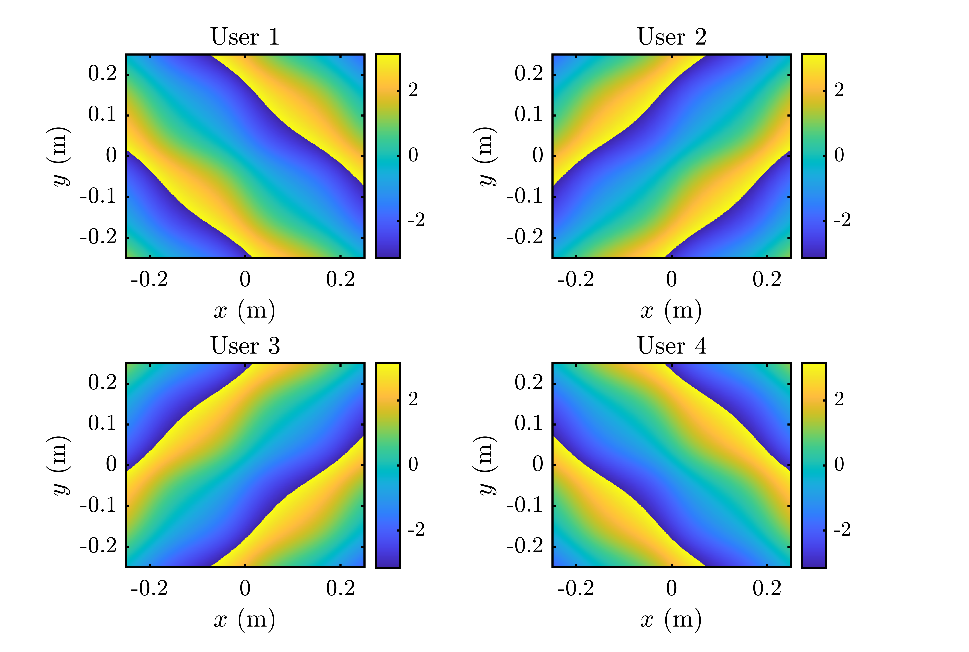}
  \caption{Phases of the source current patterns.}
\end{subfigure}
  \caption{Amplitudes and phases of the source current patterns for four users symmetrically located at $\mathbf{r}_1 = [5, 5, 15]^T$, $\mathbf{r}_2 = [5, -5, 15]^T$, $\mathbf{r}_3 = [-5, 5, 15]^T$, and $\mathbf{r}_4 = [-5, -5, 15]^T$, respectively.}
  \label{fig_pattern_1}
\end{figure} 

\begin{figure}[t!]
  \centering
  \begin{subfigure}[t]{0.5\textwidth}
  \includegraphics[width=1\textwidth]{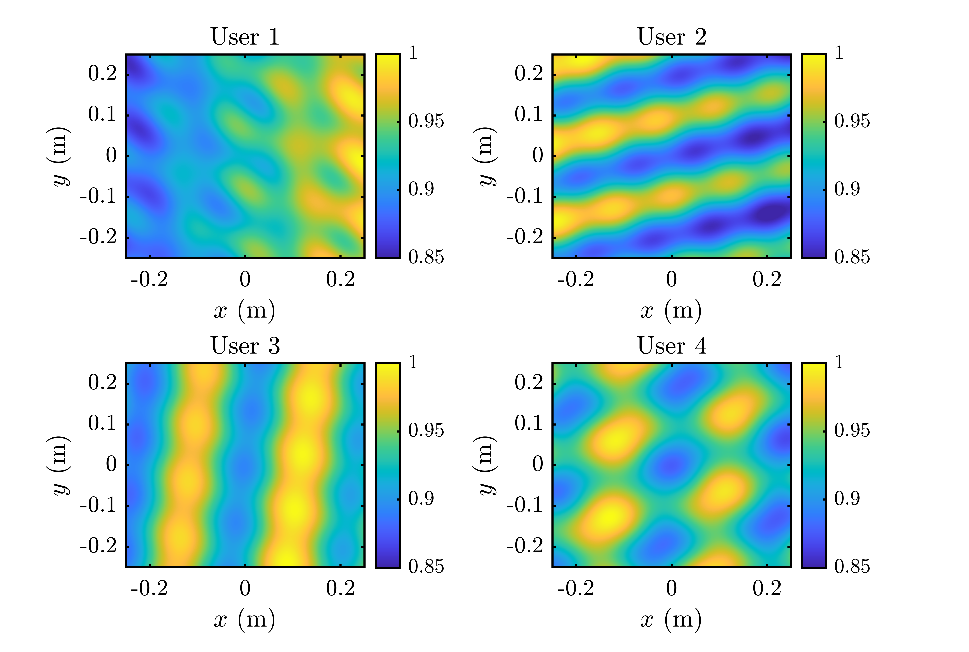}
  \caption{Amplitudes of the source current patterns.}
  \end{subfigure}
  \begin{subfigure}[t]{0.5\textwidth}
  \includegraphics[width=1\textwidth]{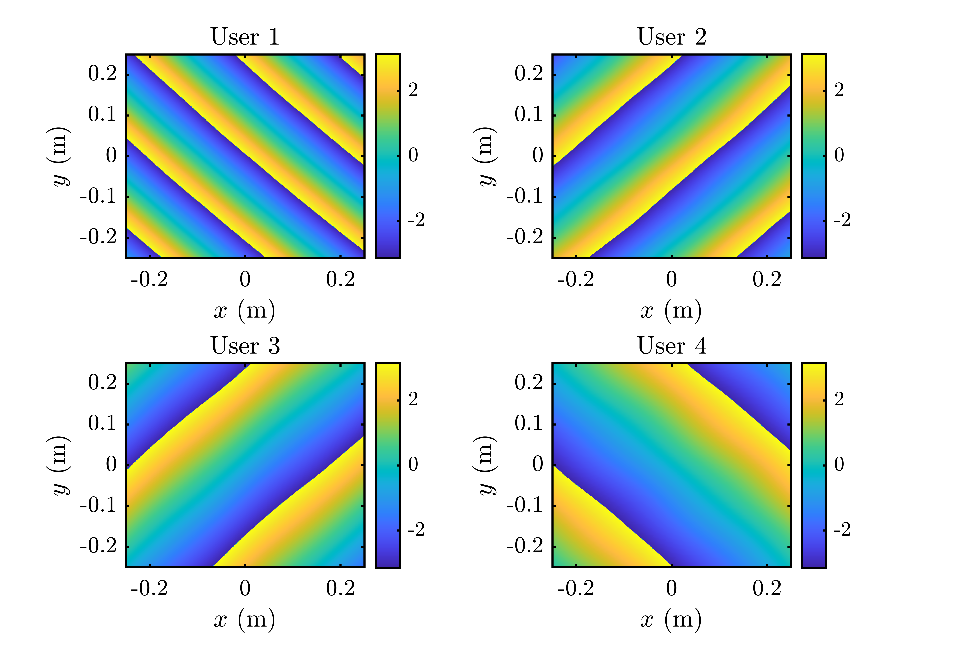}
  \caption{Phases of the source current patterns.}
\end{subfigure}
  \caption{Amplitudes and phases of the source current patterns for four users asymmetrically located at $\mathbf{r}_1 = [5, 5, 5]^T$, $\mathbf{r}_2 = [5, -5, 10]^T$, $\mathbf{r}_3 = [-5, 5, 15]^T$, and $\mathbf{r}_4 = [-5, -5, 20]^T$, respectively.}
  \label{fig_pattern_2}
\end{figure} 

\subsection{WSR Versus Number of Users}

Fig. \ref{fig_user} illustrates the impact of the number of users $K$ on the WSR performance. Recall that the weight for each user is set as $\alpha_k = 1/K, \forall k$. Under this condition, the WSR can be interpreted as the average rate across users. As expected, the proposed approaches consistently deliver the best performance compared to the benchmarks. It can be observed that the WSR achieved by all approaches decreases as the number of users increases. This phenomenon is primarily due to two factors. On the one hand, with fixed transmit power, the power allocated to each user decreases as the number of users grows. On the other hand, inter-user interference becomes stronger with more users. Furthermore, the ZF approaches exhibit a notable performance loss when the number of users is large, as the channels between users are less likely to be orthogonal in this case.

\subsection{WSR Versus Frequency}
Fig. \ref{fig_frequency} illustrates the impact of frequency $f$ on WSR performance. As with previous results, the proposed approaches consistently outperform the benchmarks across all frequencies. It is interesting to observe that a higher frequency leads to an increase in WSR, which aligns with the findings discussed in \cite{bjornson2024enabling}. This result underscores the importance of leveraging higher frequencies to boost communication capacity in future wireless systems. In this context, the advantages of the proposed approaches become more pronounced, because their computational complexity remains unaffected by frequency, in contrast to the Fourier-based approach.

\subsection{Source Current Patterns}
Fig. \ref{fig_pattern_1} shows the amplitudes and phases of the source current patterns obtained using the proposed CoV-based approach for four symmetrically located users. It is noteworthy that for symmetrically located users, the amplitudes of the source current patterns are identical for each user, while the phases are orthogonal to each other. This suggests that inter-user interference is primarily eliminated through the phase differences of the source current patterns for symmetrically located users. Fig. \ref{fig_pattern_2} further shows the source current patterns for four asymmetrically located users. In this case, it is interesting to observe that both the amplitudes and phases of the source current patterns differ significantly between users, suggesting that inter-user interference must be mitigated jointly by both amplitudes and phases.

\section{Conclusions} \label{sec:conclusion}

This paper has proposed a pair of low-complexity approaches that directly solve the beamforming optimization in multi-user CAPA systems for maximizing the WSR, namely the CoV-based approach and the Corr-ZF approach. Compared to the state-of-the-art Fourier-based approach, the proposed approaches enhance the WSR performance while significantly reducing the computational complexity at the same time. The results presented in this paper highlight the remarkable effectiveness of CoV in solving continuous integral-based optimization problems within wireless communications, revealing the potential of leveraging this powerful method to address other complex challenges in CAPA systems, such as tri-polarized CAPA beamforming, CAPA at both the transmitter and receiver, and CAPA for sensing.

\begin{appendices}

    
    \section{Proof of Lemma \ref{lemma_equal_power}} \label{lemma_equal_power_proof}
    
    Let $\{\tilde{J}_k(\mathbf{s})\}_{k=1}^K$ denote a set of feasible solution to problem \eqref{problem_se_max} that satisfies
    \begin{equation}
        \tilde{P}_{\mathrm{T}} \triangleq \sum_{k=1}^K \int_{\mathcal{S}_{\mathrm{T}}} \left|\tilde{J}_k(\mathbf{s})\right|^2 < P_{\mathrm{T}}
    \end{equation}
    By defining the scaling factor $\rho = P_{\mathrm{T}}/\tilde{P}_{\mathrm{T}}$ and the scaled solution $J_k(\mathbf{s}) = \sqrt{\rho} \tilde{J}_k(\mathbf{s})$, it can be readily shown that the SINR $\gamma_k$ in \eqref{SINR_original} achieved by the scaled solution $\{J_k(\mathbf{s})\}_{k=1}^K$ must be higher than that achieved by the original solution $\{\tilde{J}_k(\mathbf{s})\}_{k=1}^K$ due to the fact that $\rho > 1$. Additionally, it is also easy to show that 
    \begin{equation}
        \sum_{k=1}^K \int_{\mathcal{S}_{\mathrm{T}}} \left|J_k(\mathbf{s})\right|^2 = \rho \sum_{k=1}^K \int_{\mathcal{S}_{\mathrm{T}}} \left|\tilde{J}_k(\mathbf{s})\right|^2 = P_{\mathrm{T}}.
    \end{equation}
    The above results imply that for any feasible solution, a solution that achieves higher SINR and power equality constraint always exists. The proof is thus completed.
    
    \section{Proof of Lemma \ref{fundamental_lemma}} \label{fundamental_lemma_proof}
    We can employ a proof by contradiction to establish this lemma. Suppose there exists a function $V(\mathbf{s}) \neq 0$ on the set $\mathcal{S}$ such that the condition \eqref{fundamental_lemma_condition} is satisfied. Let us consider a special case of function $U(\mathbf{s})$, which is given by 
    \begin{equation}
        U(\mathbf{s}) = P(\mathbf{s}) V(\mathbf{s}).
    \end{equation} 
    Here, $P(\mathbf{s})$ is a function satisfying $P(\mathbf{s}) > 0, \forall \mathbf{s} \in \mathcal{S}$ and $P(\mathbf{s}) = 0, \forall s \in \partial \mathcal{S}$. In this case, the condition \eqref{fundamental_lemma_condition} becomes
    \begin{equation}
        \Re \left\{ \int_{\mathcal{S}} P(\mathbf{s}) |V (\mathbf{s})|^2 d \mathbf{s} \right\} = 0.
    \end{equation}
    Based on the property of $P(\mathbf{s})$ and the above condition, we must have $V (\mathbf{s}) = 0$. Therefore, we reach a contradiction. The proof is thus completed.
    
    \section{Proof of Proposition \ref{theorem_optimal_condition}} \label{theorem_optimal_condition_proof}
    
    We exploit the CoV \cite{gelfand2000calculus} to prove the \textbf{Proposition \ref{theorem_optimal_condition}}. At any local maximum of $g(J_k)$, for any $\epsilon \rightarrow 0$, we have  
    \begin{equation} \label{maximum_epsilon}
        g(J_k) \ge g(J_k + \epsilon U_k),
    \end{equation}
    where $U_k$ is any arbitrary smooth function that satisfies $U_k(\mathbf{s}) = 0, \forall \mathbf{s} \in \partial \mathcal{S}_{\mathrm{T}}$, and $\epsilon U_k$ is referred to as the variation of the function $J_k$. The right-hand side of the above inequality is a function of $\epsilon$, defined as $\Phi_k(\epsilon) \triangleq g(J_k + \epsilon U_k)$, which can be obtained according to \eqref{eqn_g_function} as 
    \begin{align} \label{variation_function}
        \Phi_k(\epsilon) = &2 \epsilon \Re \Bigg\{ A_k \int_{\mathcal{S}_{\mathrm{T}}} H_k^*(\mathbf{s}) U_k^*(\mathbf{s}) d \mathbf{s} \nonumber \\ & - \sum_{i=1}^K \Bigg( B_i \int_{\mathcal{S}_{\mathrm{T}}} \int_{\mathcal{S}_{\mathrm{T}}} H_i(\mathbf{z}) J_k (\mathbf{z}) H_i^* (\mathbf{s}) U_k^*(\mathbf{s}) d \mathbf{z} d \mathbf{s} \nonumber \\
        &+ C_i \int_{\mathcal{S}_{\mathrm{T}}} J_k(\mathbf{s}) U_k^*(\mathbf{s}) d \mathbf{s}\Bigg) \Bigg\} \nonumber \\ &+ \epsilon^2 \sum_{i=1}^K \Bigg( B_i \left| \int_{\mathcal{S}_{\mathrm{T}}} H_i(\mathbf{s}) U_k(\mathbf{s}) d \mathbf{s} \right|^2 \nonumber \\
        &+ C_i \int_{\mathcal{S}_{\mathrm{T}}} \left| U_k(\mathbf{s}) \right|^2 d \mathbf{s}\Bigg) + D_k,
    \end{align}
    where $D_k$ is a constant irrelevant to $\epsilon$. Since the functional $g(J_k)$ attains its local maximum at $J_k$, the function $\Phi_k(\epsilon)$ must have its maximum at $\epsilon = 0$ according to \eqref{maximum_epsilon}, leading to the following condition 
    \begin{equation}
        \left. \frac{d \Phi_k(\epsilon)}{d \epsilon} \right|_{\epsilon = 0} = 0.
    \end{equation}   
    According to \eqref{variation_function}, the above condition can be rewritten as 
    \begin{align} \label{optimal_condition_0}
        \Re \left\{\int_{\mathcal{S}_{\mathrm{T}}} U_k^*(\mathbf{s}) V_k(\mathbf{s}) d \mathbf{s}\right\} = 0,
    \end{align}
    where 
    \begin{align}
        V_k(\mathbf{s}) =  A_k H_k^*(\mathbf{s}) - \sum_{i=1}^K B_i H_i^*(\mathbf{s}) \int_{\mathcal{S}_{\mathrm{T}}} H_i (\mathbf{z}) J_k (\mathbf{z}) d \mathbf{z} \nonumber \\ 
        -\sum_{i=1}^K C_i J_k (\mathbf{s}).
    \end{align}
    Note that the condition \eqref{optimal_condition_0} must be satisfied for any arbitrary function $U_k$. Therefore, according to \textbf{Lemma \ref{fundamental_lemma}}, we must have 
    \begin{equation}
        V_k(\mathbf{s}) = 0, \forall \mathbf{s} \in \mathcal{S}_{\mathrm{T}},
    \end{equation}
    which can be reformulated as \eqref{optimal_condition}. The proof is thus completed.
    
    

    \end{appendices}

\balance
\bibliographystyle{IEEEtran}
\bibliography{reference/mybib}

\end{document}